\documentclass[preprintnumbers,secnumarabic,amsmath,amssymb,nofootinbib,floatfix]{revtex4}
\usepackage{varioref,exscale,latexsym,amsmath,amssymb}
\usepackage{graphicx,epstopdf}

\usepackage{epsfig}
\usepackage{graphicx}
\usepackage{dcolumn}
\usepackage{bm}
\usepackage{simplewick}

\def\cN{\mathcal{N}}
\def\cM{\mathcal{M}}

\def\cMt{\overset{_0}{\mathcal{M}}}

\def\sq[#1,#2]{\left[#1\,#2\right]}
\def\an[#1,#2]{\left\langle#1\,#2\right\rangle}
\def\spab[#1,#2,#3]{\left\langle#1|#2|#3\right]}

\def\beq{\begin{equation}}

\def\eeq{\end{equation}}

\def\beqa{\begin{eqnarray}}

\def\eeqa{\end{eqnarray}}

\begin{document}

\title{{\bf Low Energy Theorems of Quantum Gravity from Effective Field Theory }}

\medskip\
{\author{ John F. Donoghue}
\author{Barry R. Holstein}
\affiliation{Department of Physics,
University of Massachusetts\\
Amherst, MA  01003, USA}

\begin{abstract}
In this survey, we review some of the low energy quantum predictions of General Relativity which are independent of details of the yet unknown high-energy completion of the gravitational interaction. Such predictions can be extracted using the techniques of effective field theory.
\end{abstract}
\maketitle
\section{Introduction}

The study of quantum gravity is a major research field. Over the past two decades, there has been a transition in the understanding of this subject that has not yet been fully absorbed by scientists outside of the field. It used to be stated that general relativity and quantum mechanics were incompatible. There were many reasons given for this conflict, some of which look foolish from the modern perspective. However, a modern view is that general relativity forms a quantum effective field theory at low energies. As described below, effective field theory is a standard technique to describe quantum effects at low energy where one knows the active degrees of freedom and their interactions. The effective field theory allows predictions which are valid at those energies. This does not eliminate the need to understand gravity at very high energies where many interesting effects occur. However it is still remarkable progress, as we now understand that gravity and quantum mechanics can be compatible at the energies that have been experimentally probed.

The effective field theory treatment allows the separation of quantum effects which follow from known low energy physics from those that depend on the ultimate high energy completion of the theory of gravity. The key ingredient follows from the uncertainty principle in that high energy effects are very local while those from low energy are non-local. Indeed there are some results that can be described as ``low energy theorems'' of quantum gravity. This means that they are the outcome of {\em any} consistent theory of quantum gravity. The only assumptions of that full quantum gravity theory is that it limits to general relativity at low energy. Given how much the combination of gravity and quantum mechanics has been maligned in the past, it is remarkable that such universal results for quantum gravity can now be formulated.

In this review, we provide a survey of some of the low energy theorems which have been calculated thus far. Most of these are scattering amplitudes, as these are the structures that quantum field theory calculates most readily. In the future we hope that equivalent results can be developed for other gravitations settings.

\section{Effective Field Theory}

Physics is an experimental science. We only reliably know the degrees of freedom and their interactions which have been probed by present experiment. This means that there almost certainly exists new physics to be found beyond the present energies. For gravity, this caveat is especially relevant since the Planck scale of $M_P =G^{-{1\over 2}}=1.2 \times 10^{19}$~GeV, where $G$ is the Cavendish constant, seems to be the obvious location for new physics. Experimentally, however, we are so far from the Planck scale that we have little hope of uncovering the nature of this new physics in the foreseeable future.

Quantum mechanics does seem to care about this unknown new physics, since in perturbative calculations we are instructed to sum over a complete set of intermediate states---at {\it all} energies---when making quantum calculations.  This difficulty is solved, however, by a simple mechanism---the Heisenberg uncertainty principle.  The effects of very high energy appear as short-distance phenomena to us and thus appear as {\it local} terms in a Lagrangian.  The coefficients of these local terms are the residual manifestation of the high energy physics.  Most of these Lagrangians are basically irrelevant because they are suppressed by powers of the heavy scale.  Thus the unknown physics is reduced to a very few parameters which we can either measure (if we do not know the high energy theory) or predict (if we think that we do know this theory).  

On the other hand, low energy physics is of a very different character, since light particles can propagate long distances and their low energy effects are not local.  The distinction between local and non-local is the key to separating the physics of low energy from high energy.  To be sure, even light particles, when treated in loop diagrams, can have effects from very high energies since loops integrate over all energies.  These high energy effects are not reliable, as we have not yet experimentally probed that part of the theory.  But this is not a problem.  Again these high energy effects are local, and merely are absorbed into the renormalization of the coefficients of the local effective Lagrangian.

Effective field theory (EFT) is the technique that takes advantage of these physical properties in order to make predictions at low energies which are reliable, because they utilize only the low energy particles and their low energy interactions. Useful references on EFT and its applications can be found in \cite{std, efr}.  

As a more technical explanation, consider a general quantum field theory with light and heavy degrees of freedom.  In quantum field theory it is straightforward to represent what is going on
by use of path integral methods.  If $\phi,\,(\Phi)$ represents a light (heavy) field respectively, then the
functional integral which characterizes the full quantized theory is given by (note we are using $\hbar=1$
\begin{equation}
W=\int[d\phi][d\Phi]\exp\, i\int d^4x{\cal L}(\phi,\Phi).
\end{equation}
Now suppose that we integrate out the heavy degrees of freedom.  What is left is a functional integral in terms of a non-local ``effective'' interaction which characterizes the theory in terms of only the light degree of freedom $\phi$
\begin{equation}
W=N\int[d\phi]\exp i\int d^4x{\cal L}_{eff}(\phi)
\end{equation}
but which includes the {\it virtual} effects of the heavy degrees of freedom $\Phi$ to all orders.

In this setting, it is possible to produce low energy theorems.  This phrase refers to calculations of amplitudes or relations between amplitudes that remain valid independent of any modification of the high energy component of the theory.  This can only happen if such relations depend uniquely on the low energy part of the theory.  {\it Any} high energy theory that is capable of generating these particles and these interactions at low energy must yield identical results.  Effective field theory allows us to calculate these relations.  In the context of this review, the conditions for the existence of these low energy theorems is that the ultimate quantum gravity theory must reduce to general relativity in four dimensions at low energy.  Conventional quantum field theory, expressed through path integral quantization, is assumed to apply.  There are not really any other ingredients.  This combination of general relativity and quantum field theory automatically behaves as an effective field theory at the lowest energies, and we can extract low energy theorems by the use of effective field theory, as shown below.

\section{Low Energy Theorems of QCD}

In order to understand the use of effective theory in a non-renormalizable theory such as general relativity, it is useful to first examine low energy quantum chromodynamics (QCD).  In QCD at the lowest energies there exist only light pions which are dynamically active and the interactions of these pions are constrained by the original chiral symmetery of QCD\cite{std, efr, gle}. The resulting effective
field theory---chiral perturbation theory---has many aspects in common with general relativity.  Chiral perturbation theory has been exceptionally well studied both theoretically and experimentally.  We provide a somewhat detailed review here in order to set the stage for a parallel treatment of general relativity.

In order to understand low energy QCD, we begin by introducing the property of ``chirality'', defined by the operators
\begin{equation}
\Gamma_{L,R} = {1\over 2}(1\pm\gamma_5)={1\over 2}
\left( \begin{array}{c c }
1 & \mp 1 \\
\mp 1 & 1
\end{array}\right)
\end{equation}
which project out left- and right-handed components of the Dirac wavefunction of quarks
via
\begin{equation}
\psi_L = \Gamma_L \psi \qquad \psi_R=\Gamma_R
\psi \quad\mbox{with}\quad \psi=\psi_L+\psi_R
\end{equation}
In terms of these chirality states the light ($u,d$) quark component of the QCD Lagrangian
can be written as
\begin{equation}
{\cal L}_{QCD}^{ud}=\bar{q}(i\not\! \! D-m)q=\bar{q}_Li\not \! \! D q_L +
\bar{q}_Ri\not\!\! D q_R -\bar{q}_L m q_R-\bar{q}_R m q_L
\end{equation}
where $q=\left(\begin{array}c u\\d\end{array}\right)$ is a two component spinor and $m=(m_u,m_d)_{diag}$ is the $u,d$ quark mass matrix.  We note that in the limit of vanishing mass
\begin{equation}
{\cal L}_{\rm QCD}^{ud}\underset{m\rightarrow 0}{\longrightarrow}\bar{q}_L i \not\!\! D q_L +
\bar{q}_R i \not\!\! D q_R
\end{equation}
is invariant under {\it independent} global left- and right-handed rotations
\begin{equation}
q_L  \rightarrow \exp (i \sum_{j=1}^3 \tau_j\alpha_j)
q_L\equiv Lq_L,\qquad
q_R  \rightarrow \exp (i\sum_{j=1}^3 \tau_j \beta_j)
q_R\equiv Rq_R
\end{equation}
This is
$SU(2)_L \bigotimes SU(2)_R$ invariance or chiral $SU(2)\bigotimes SU(2)$.  Continuing
to neglect the light quark masses, we see that in a chiral symmetric world one might expect six---three
left-handed and three right-handed---conserved Noether currents
\begin{equation} \bar{q}_L\gamma_{\mu} {1\over 2} \tau_i q_L \, ,
\qquad \bar{q}_R\gamma_{\mu}{1\over 2}\tau_i
q_R \end{equation}
Equivalently, by taking the sum and difference of the left- and right-handed currents we should have three conserved polar vector and
three conserved axial vector currents
\begin{equation}
V^i_{\mu}=\bar{q}\gamma_{\mu} {1\over 2}\tau_i q,\qquad
A^i_{\mu}=\bar{q}\gamma_{\mu}\gamma_5{1\over 2} \tau_i q
\end{equation}

The polar vector symmetry is seen in the particle spectrum and is just isospin symmetry. However, the axial symmetry is not observed in a Wigner-Weyl fashion---there are {\it no} parity doublets---but rather is spontaneously broken.  The implication of this breaking is that, via Goldstone's theorem\cite{gst}, there must exist nearly massless pseudoscalar particles--- pions---which are approximate Goldstone bosons for the chiral symmetry of QCD.  Moreover, the pion interactions must obey numerous symmetry restrictions.  The simplest way to keep track of these symmetry requirements is to write a general effective Lagrangian which obeys this symmetry.  This can be accomplished by use of a nonlinear function of the pion field,
\begin{equation}
U = \exp \left(i \frac{\boldsymbol{\tau}\cdot\boldsymbol{\pi}}{F_\pi}\right)~~~\mbox{transforming as}~~~U\to LUR^\dagger
\end{equation}
with $L,~R$ being the $2\times 2$ $SU(2)$ matrices in $SU(2)_{L,R}$ respectively which were introduced above.  Lagrangians constructed using $U$ retain the chiral symmetry and obey the symmetry requirements. The general Lagrangian is then constructed in an expansion in the number of derivatives, so that the terms with the fewest derivatives will be most important at low energy.

We infer then that the {\it lowest order} SU(2) effective chiral Lagrangian
can be written as
\begin{equation}
 {\cal L}^{(2)}={F_\pi^2 \over 4} \mbox{Tr} (\partial_{\mu}U \partial^{\mu}
 U^{\dagger})-{m^2_{\pi}\over 4} F_\pi^2 \mbox{Tr} (U+U^{\dagger})\,  .\label{eq:abc}
\end{equation}
where the superscript 2 indicates that we are working at two-derivative order
or one power of chiral symmetry breaking---{\it i.e.} $m_\pi^2\propto m_u+m_d$.
This Lagrangian is also {\it unique}---if we expand to order $\phi^2$
\begin{equation}
\mbox{Tr}\partial_{\mu} U \partial^{\mu} U^{\dagger} =
\mbox{Tr} {i\over F_\pi} \boldsymbol{\tau}\cdot\partial_{\mu}\boldsymbol{\phi} \times
{-i\over F_\pi}\boldsymbol{\tau}\cdot\partial^{\mu}\boldsymbol{\phi}= {2\over F_\pi^2}
\partial_{\mu}\boldsymbol{\phi}\cdot \partial^{\mu}\boldsymbol{\phi}\,  ,
\end{equation}
we reproduce the free pion Lagrangian, as required---
\begin{equation}
 {\cal L}^{(2)}_{\phi^2} ={1\over 2} \partial_{\mu}
\boldsymbol{\phi}\cdot \partial^{\mu} \boldsymbol{\phi} -{1\over 2} m^2_{\pi}
\boldsymbol{\phi}\cdot \boldsymbol{\phi} \,  .
\end{equation}

At higher orders interactions are generated.  For example, in the case of pion scattering,
expanding ${\cal L}^{(2)}$ to order $\phi^4$ we find
\begin{equation}
{\cal L}^{(2)}_{\phi^4}={1\over 6F_\pi^2} \boldsymbol{\phi}\cdot\boldsymbol{\phi}\boldsymbol{\phi}\cdot \Box\boldsymbol{\phi} +
{1\over 2F_\pi^2} (\boldsymbol{\phi}\cdot\partial_{\mu}\boldsymbol{\phi})^2
 + { m^2_\pi \over 24 F_\pi^2}(\boldsymbol{\phi}\cdot\boldsymbol{\phi})^2
\end{equation}
which yields for the on shell $\pi\pi$ $T$-matrix
\begin{equation}
T_{ab;cd}(q_a, q_b;q_c, q_d)  = {1\over F_{\pi}^2} \left[ \delta_{ab}
\delta_{cd}(s-m^2_{\pi})+\delta_{ab}\delta_{bd}(t-m_{\pi}^2)
+\delta_{ad}\delta_{bc}(u -m_{\pi}^2)\right]\label{eq:uh}
\end{equation}
Defining more generally
\begin{equation}
 T_{ab;cd}(s,t,u)=A(s,t,u)\delta_{ab}\delta_{cd}
 +A(t,s,u)\delta_{ac}\delta_{bd}
 +A(u,t,s)\delta_{ad}\delta_{bc} \, ,
\end{equation}
we can write the chiral prediction in terms of the more conventional isospin
language by taking appropriate linear combinations
\begin{eqnarray}
T^0(s,t,u)&=& 3A(s,t,u)+A(t,s,u)+A(u,t,s)\, , \nonumber \\
T^1(s,t,u)&=& A(t,s,u)-A(u,t,s)\, , \nonumber \\
T^2(s,t,u)&=& A(t,s,u)+A(u,t,s)\, .
\end{eqnarray}
Partial wave amplitudes, projected out via
\begin{equation}
T_l^I(s)={1\over 64\pi}\int^1_{-1} d(\cos\theta) P_l(\cos \theta)
T^I(s,t,u)  \, ,
\end{equation}
can be used to identify the associated scattering phase shifts via
\begin{equation}
 T^I_l(s) = \left( {s\over s-4m^2_{\pi}}\right)^{1\over 2} e^{i\delta^I_l}
\sin \delta^I_l \,  .
\end{equation}
Then from the lowest order chiral form Eq. (\ref{eq:uh})
\begin{equation}
A(s,t,u)={s-m^2_{\pi}\over F^2_{\pi}}
\end{equation}
we generate predictions for the pion scattering lengths and effective ranges\cite{wei}
\begin{eqnarray}
a^0_0&=&{7m^2_{\pi}\over 32\pi F^2_{\pi}}\, , \quad
a^2_0=-{m^2_{\pi}\over 16\pi F^2_{\pi}}\, , \quad
a^1_1=-{m^2_{\pi}\over 24\pi F^2_{\pi}}\, ,  \nonumber \\
b^0_0&=&{m^2_{\pi}\over 4\pi F^2_{\pi}}\, , \quad
b^2_0={m^2_{\pi}\over 8\pi F^2_{\pi}}\, ,
\end{eqnarray}
comparison of which with experimental numbers is shown in Table 1. These are the start of some low energy theorems of QCD\cite{cga}.
\begin{table}
\begin{center}
\begin{tabular}{l r r r}
\hline \hline
& Experimental & Lowest Order&
First Two Orders\\ \hline
$a^0_0$ & $0.220\pm  0.005 $& 0.16 & 0.20 \\
$b^0_0$ &$ 0.250\pm  0.030$ & 0.18 & 0.26 \\
$a^2_0$ &$-0.044 \pm  0.001$ & -0.045 &-0.041 \\
$b^2_2$ &$-0.082 \pm  0.008$ & -0.089 &-0.070 \\
$a^1_1$ &$0.038 \pm  0.002$ & 0.030 &0.036 \\
$b^1_1$ & 0 & 0.043 \\
$a^0_2$ & $(17\pm 3) \times 10^{-4 }$& 0 & $20 \times 10^{-4}$ \\
$a^2_2$ & $(1.3\pm 3) \times 10^{-4 }$& 0 & $3.5 \times 10^{-4}$ \\
\hline\hline
\end{tabular}
\caption{The pion scattering lengths and slopes compared with
predictions of chiral symmetry. The last column has been taken from \cite{Colangelo}}
\end{center}
\end{table}

However, we can do better by considering loop effects. These will generate corrections to the tree amplitudes and will
bring in imaginary parts for the amplitude that are necessary to satisfy unitarity. We will see that these corrections are
expressed in an expansion in the energy, such that the tree results are the lowest energy results and one loop results are corrections to the tree level. Inclusion of loop effects comes with a price---numerous divergences are introduced and this difficulty prevented progress in this field for nearly a decade\cite{ggf} until a paper by Weinberg suggested the solution\cite{wbr}---dealing with such divergences,
just as in QED, by introducing phenomenologically determined counterterms into the Lagrangian in order to absorb the infinities.  We show in the next section how this can be accomplished.

\subsection{Effective Chiral Lagrangian for QCD}

We now apply Weinberg's suggestion to the effective chiral Lagrangian, Eq.
(\ref{eq:abc}).  As noted above, when loop corrections are made to lowest order
amplitudes in order to enforce unitarity, divergences inevitably arise.
However, there is an important difference from the familiar case of QED in that the
form of the divergences is {\it different} from their lower order
counterparts---the theory is nonrenormalizable!
The reason for this can be seen from a simple example---$\pi\pi$ scattering.  In lowest order there exists a
tree level contribution from ${\cal L}^{(2)}$ which is ${\cal O} (p^2/F_\pi^2)$ or ${\cal O}(m_\pi^2/F_\pi^2)$
where $p$ represents some generic external energy-momentum.  The fact that
$p$ appears to the second power is due to the feature that its origin is
the {\it two}-derivative Lagrangian ${\cal L}_2$.  Now suppose that $\pi\pi$
scattering is examined at one loop order.  Since the scattering amplitude
must still be dimensionless but now the amplitude involves a factor
$1/F_\pi^4$ the numerator must involve {\it four} powers of energy-momentum or two powers of energy-momentum together with $m_\pi^2$
or $m_\pi^4$.  Thus any counterterm which is included in order to absorb this divergence
must be {\it four}-derivative in character.  Gasser and Leutwyler studied
this problem and wrote the most general form of such an ${\cal O}(p^4)$
counterterm in chiral SU(2) as\cite{gle}

\begin{eqnarray}
{\cal L}_4 &  =&\sum^{7}_{i=1} \ell_i {\cal O}_i
= {1\over 4}\ell_1\bigg[{\rm tr}(D_{\mu}UD^{\mu}U^{\dagger})
\bigg]^2+{1\over 4}\ell_2{\rm tr} (D_{\mu}UD_{\nu}U^{\dagger})\cdot
{\rm tr} (D^{\mu}UD^{\nu}U^{\dagger}) \nonumber \\
&+&{1\over 16}\ell_3\bigg[ {\rm tr} \left(\chi U^{\dagger}+
U \chi^{\dagger}\right)\bigg]^2+{1\over 4}\ell_4{\rm tr}\left[D_\mu U^\dagger D^\mu\chi+D_\mu U D^\mu\chi^\dagger\right]\nonumber\\
&+&{1\over 2}\ell_5{\rm tr}\left(F^L_{\mu\nu}
U F^{R\mu\nu}U^{\dagger}\right)+{i\over 2}\ell_6{\rm tr} \left(F^L_{\mu\nu}D^{\mu}U D^{\nu}
U^{\dagger}+F^R_{\mu\nu}D^{\mu} U^{\dagger}
D^{\nu} U \right)\nonumber\\
&+&{1\over 16}\ell_7\bigg[ {\rm tr} \left(\chi^{\dagger}U-
U\chi^{\dagger}\right)\bigg]^2
\end{eqnarray}
where the covariant derivative is defined via
\begin{equation}
D_\mu U=\partial_\mu U+\{A_\mu,U\}+[V_\mu,U],
\end{equation}
the constants $\ell_i, i=1,2,\ldots 7$ are arbitrary (not determined from chiral
symmetry) and
$F^L_{\mu\nu}, F^R_{\mu\nu}$ are external field strength tensors defined via
\begin{eqnarray}
F^{L,R}_{\mu\nu}=\partial_\mu F^{L,R}_\nu-\partial_\nu
F^{L,R}_\mu-i[F^{L,R}_\mu ,F^{L,R}_\nu],\qquad F^{L,R}_\mu =V_\mu\pm A_\mu .
\end{eqnarray}
Now just as in the case of QED, the bare parameters $\ell_i$ which appear
in this Lagrangian are {\it not} physical quantities.  Instead the experimentally
relevant (renormalized)
values of these parameters are obtained by appending to these bare values
divergent one-loop contributions having the form
\begin{equation}
 \ell^r_i = \ell_i +{\gamma_i\over 32\pi^2}
\left[{1\over \epsilon}-\gamma +\ln (4\pi)+1\right]
\end{equation}
where $\gamma_i$ are calculable constants\cite{gle} and, evaluating the loop integrals in $d$ dimensions, $\epsilon=(d-4)/2$ and $\gamma$ is the Euler constant.
By comparing with experiment, Gasser and Leutwyler were able to determine empirical
values for the seven $\ell_i^r$.   While seven sounds like a rather large number,
this picture is actually quite predictive\cite{csr, crv}.

\subsection{Full effective field theory for QCD}

The use of effective Lagrangians is not the full content of an effective field theory. These do express some symmetry restrictions on amplitudes and also describe the residual effects from high energy. But effective field theory is a full quantum field theory with trees and loops of the particles that are described in the Lagrangian. The renormalization described above is only a part of the effects of loops, and not even an interesting part of the loops. Much more relevant for physics are the low energy effects of the loops.

The actual calculations are straightforward and in the results we see several effects. There are the divergences described above, and these are absorbed into the renormalized parameters of the Lagrangian in a specified renomrmalization scheme. Then there are these parameters themselves. Here we are faced with various possibilities. We can measure the parameters in other reactions, or perhaps turn to lattice calculations in order to predict them from the full theory of QCD. These numbers themselves are not predictions of the effective field theory. The distinction here is that the effective field theory has its structure determined by the symmetry and light degrees of freedom only, and can serve as the low energy limit of any theory which has these features. For example, the linear sigma model and also QCD-like theories with different numbers of colors all have the same structure for the effective field theory. However, the values of the constants in the Lagrangian would differ for such theories. In this sense, the constants encode the physics of the ultimate high energy theory (often referred to as the ``UV completion'' of the EFT). Because these terms are local, this is what is expected from the uncertainty principle arguments discussed earlier. 
 
So what are the quantum predictions? We have argued that they are not the renormalization procedure nor the parameters themselves. However, there are residual quantum effects that are independent of the renormalization and of the parameters. These come from low energy propagation of the light particles. (The high energy parts of loops are local and respect the symmetry, so they contribute shifts to the renormalized parameters.) In advance we can know some things to look for. The local chiral lagrangian is expressed in powers of the derivatives. The resulting amplitudes in momentum space are then polynomials in the energies involved - these are an analytic expansion of the amplitude. However loops also bring in non-analytic terms, such as $\ln ( -q^2)$  or $\sqrt{-q^2}$, which cannot arise from an expansion of a local Lagrangian. These non- analytic terms are signals of long distance propagation by light particles. An important byproduct is that such non-analytic terms at one loop are always independent of any of the parameters of the Lagrangian as well as being finite and divergence free. 

In pure pion physics one finds uniquely the logarithmic non-analyticity, as this arises in the bubble diagrams that appear in the pionic theory. The square root $\sqrt{-q^2}$ arises in triangle diagrams with one massive and two massless particles, and then appears in pion interactions with baryons. In real QCD with massive pions, these non-analytic terms also appear as $\ln ( m_\pi^2)$  or $\sqrt{m_\pi^2}$. Such mass dependence is also a unique prediction of the effective field theory.

The results in the third column of Table I are the results of this program carried out to two loop order\cite{Colangelo} also taking into account dispersion relations constraints, which is probably the gold standard for the pionic effective field theory of QCD.  However there is an extensive literature of one loop studies of many processes. The full review of this program is not appropriate for the present document, but we should note that the program has been quite successful. We refer the reader to the literature for further information.

\section{Effective Field Theory of Gravity}

Having seen how the strong interactions can be described via effective field theoretic methods, we move on to our primary goal, which is to treat the gravitational interaction in a parallel fashion, as developed in refs. \cite{jfd}.  Reviews of this procedure can be found in \cite{cbg, jfd2}. The EFT method relies on an expansion in energy-momentum or derivatives and the gravitational interaction is described in this way.  In order to understand how this is done we require the connection coefficient $\Gamma^\lambda_{\alpha\beta}$, which is defined via the covariant derivative $D_\alpha$ in terms of its operation on a vector field $A^\lambda$ as
\begin{equation}
D_\alpha A^\lambda=\partial_\alpha A^\lambda+\Gamma^\lambda_{\alpha\beta}A^\beta
\end{equation}
where, in terms of the metric tensor $g_{\mu\nu}$,
\begin{equation}
\Gamma^\lambda_{\alpha\beta}={1\over 2}g^{\lambda\zeta}\left[\partial_\alpha g_{\beta\zeta}+\partial_\beta g_{\alpha\zeta}-\partial_\zeta g_{\alpha\beta}\right].
\end{equation}
Taking the metric tensor $g_{\mu\nu}$ as a field, we see that the connection involves a single field derivative, while the curvature tensor
\begin{equation}
R^\alpha_{\mu\nu\beta}=\partial_\nu\Gamma^\alpha_{\mu\beta}-\partial_\beta\Gamma^\alpha_{\mu\nu}+
\Gamma^\zeta_{\mu\beta}\Gamma^\alpha_{\zeta\nu}-\Gamma^\zeta_{\mu\nu}\Gamma^\alpha_{\zeta\beta}
\end{equation}
defined as
\begin{equation}
[D_\mu ,D_\nu]A_\alpha\equiv R^\beta_{\alpha\mu\nu}A_\beta
\end{equation}
has two, as do its associated quantities, the Ricci tensor
\begin{equation}
R_{\mu\nu}=\partial_\nu\Gamma^\alpha_{\mu\alpha}-\partial_\alpha\Gamma^\alpha_{\mu\nu}+
\Gamma^\zeta_{\mu\alpha}\Gamma^\alpha_{\zeta\nu}-\Gamma^\zeta_{\mu\nu}\Gamma^\alpha_{\zeta\alpha}
\end{equation}
and the scalar curvature
\begin{equation}
R=g^{\mu\nu}R_{\mu\nu}.
\end{equation}
The gravitational action can then be written as a derivative expansion
\begin{equation}
S_{grav}=\int d^4x\sqrt{-g}\left[\Lambda+{2\over \kappa^2}R+c_1R^2+c_2R_{\mu\nu}R^{\mu\nu}+\ldots\right]\label{eq:pz}
\end{equation}
where $\Lambda$ is the cosmological constant and $\kappa$ is given in terms of the Cavendish constant $G$ via $\kappa^2=32\pi G$.  The expansion parameter here is the Planck mass $M_P=G^{-{1\over 2}}\sim 10^{19}$ GeV.  Thus the higher order terms in the derivative expansion are suppressed by powers of $E/M_P,p/M_P<<<1$ so that the Einstein action should provide an extremely precise picture of gravitational effects at presently relevant energies\cite{hdo}.  Of course, a mystery is why the leading term in the expansion, the cosmological constant, is so small experimentally---$\Lambda\sim 10^{-47}$ GeV$^4$\cite{wbb}. (Note that a corresponding situation exists in QCD, where the smallness of the experimental result for the theta-term---$\theta\sim 10^{-{11}}$---remains unexplained\cite{std}.)  However, we shall merely take this result as an empirical fact and will neglect $\Lambda$ for the remainder of this paper.  The full action is then given by including the matter Lagrangian ${\cal L}_{mat}$.  Varying the lowest order action
\begin{equation}
S_{tot}=S_{grav}+S_{mat}=\int d^4x\sqrt{-g}\left({2\over \kappa^2}R+{\cal L}_{mat}\right)
\end{equation}
we find the Einstein equation
\begin{equation}
R_{\mu\nu}-{1\over 2}g_{\mu\nu}R=-8\pi GT_{\mu\nu}
\end{equation}
where
\begin{equation}
T_{\mu\nu}=-{2\over \sqrt{-g}}{\partial\over \partial g_{\mu\nu}}{\cal L}_{mat}
\end{equation}
is the energy-momentum tensor of matter\cite{wbb}.

The theory can be quantized using the background field method by defining
\begin{equation}
g_{\mu\nu}(x)=\bar{g}_{\mu\nu}(x)+ \kappa h_{\mu\nu}(x)
\end{equation}
where $\bar{g}_{\mu\nu}(x)$ is a classical solution of the Einstein equation.  The inverse metric tensor is then given by
\begin{equation}
g^{\mu\nu}(x)=\bar{g}^{\mu\nu}- \kappa h^{\mu\nu}(x)+ \kappa^2 h^{\mu\zeta}{h_\zeta}^\nu+\ldots
\end{equation}
where indices are raised and lowered with the classical metric tensor $\bar{g}_{\mu\nu}$.  We choose to quantize about flat space, so that
$\bar{g}_{\mu\nu}=\eta_{\mu\nu}$.  We find then for the Ricci tensor and scalar curvature at one derivative order
\begin{eqnarray}
R^{(1)}_{\mu\nu}&=&{\kappa\over 2}\left[\partial_\mu\partial_\nu h+\Box h_{\mu\nu} -\partial_\mu\partial_\zeta {h^\zeta}_\nu-\partial_{\nu}\partial_{\zeta}{h^\zeta}_\mu\right]\nonumber\\
&=&{\kappa\over 2}\left[\Box h_{\mu\nu}-\partial_\mu\left((\partial\cdot h)_\nu-{1\over 2}\partial_\nu h\right)-\partial_\nu\left((\partial\cdot h)_\mu-{1\over 2}\partial_\mu h\right)\right]\nonumber\\
R^{(1)}&=&\kappa \left[\Box h-\partial_\mu\partial_\nu h^{\mu\nu}\right]\nonumber\\
&=&{\kappa\over 2}\left[\Box h-2\partial^\sigma\left((\partial\cdot h)_\sigma-{1\over 2}\partial_\sigma h\right)\right]
\end{eqnarray}
where $h=\eta^{\mu\nu}h_{\mu\nu}$, $(\partial\cdot h)_\sigma=\partial^\lambda h_{\lambda\sigma}$, and $\Box=\eta^{\mu\nu}\partial_\mu\partial_\nu$.  As is well known the Einstein
equations are invariant under a general coordinate transformation which, in terms of the fields, implies a gauge invariance
\begin{eqnarray}
x^\mu\rightarrow x^{'\mu}&=&x^\mu+\epsilon^\mu(x)\nonumber\\
h_{\mu\nu}(x)\rightarrow {h'}_{\mu\nu}(x)&=&h_{\mu\nu}(x)-\partial_\mu\epsilon_\nu(x)-\partial_\nu\epsilon_\mu(x)
\end{eqnarray}
for infinitesimal $\epsilon^\mu(x)$.  In order to deal with this invariance, we must make a gauge choice and we
elect to work in harmonic or deDonder gauge---$g^{\mu\nu}
\Gamma^\lambda_{\mu\nu}=0$---which reads, to first order in the field
expansion,
\begin{equation}
0=\partial^\beta h_{\beta\alpha}-{1\over 2}\partial_\alpha h=(\partial\cdot h)_\alpha-{1\over 2}\partial_\alpha h\label{eq:tr}
\end{equation}
whereby the linearized Einstein equation
\begin{eqnarray}
&&R^{(1)}_{\mu\nu}-{1\over 2}\eta_{\mu\nu}R^{(1)}={\kappa\over 2}\left[\Box(h_{\mu\nu}-{1\over 2}\eta_{\mu\nu}h)
-\partial_\mu\left((\partial\cdot h)_\nu-{1\over 2}\partial_\nu h\right)\right.\nonumber\\
&-&\left.\partial_\nu\left((\partial\cdot h)_\mu-{1\over 2}\partial_\mu h\right)+\eta_{\mu\nu}\partial^\alpha\left((\partial\cdot h)_\alpha-{1\over 2}\partial_\alpha h\right)\right]=-{1\over 4}\kappa^2 T_{\mu\nu}^{mat}\nonumber\\
\quad\label{eq:jq}
\end{eqnarray}
becomes
\begin{equation}
\Box\left(h_{\mu\nu}-{1\over 2}\eta_{\mu\nu}h\right)=-{1\over 2}\kappa T^{mat}_{\mu\nu}\label{eq:fx}
\end{equation}
or its equivalent form
\begin{equation}
\Box h_{\mu\nu}=-{1\over 2}\kappa \left(T^{mat}_{\mu\nu}-{1\over 2}\eta_{\mu\nu}T^{mat}\right),\label{eq:td}
\end{equation}
where $T^{mat}=\eta^{\mu\nu}T^{mat}_{\mu\nu}$.  There exist, of course, well known solutions to Eq. (\ref{eq:td}).  For example, in the case of a stationary point mass $m$ located at the origin we have $T^{mat}_{\mu\nu}(x)=\eta_{\mu 0}\eta_{\nu 0}m\delta^3(x)$, for which the solution of Eq. (\ref{eq:td}) is
\begin{equation}
h^{(1)}_{\mu\nu}=\delta_{\mu\nu}f(r)\label{eq:cv}
\end{equation}
with
\begin{equation}
f(r)=-\kappa{m\over 16\pi r}=-\sqrt{G\over 8\pi}{m\over r}.
\end{equation}

At next order it is useful to write the Einstein equation as
\begin{equation}
R^{(1)}_{\mu\nu}-{1\over 2}\eta_{\mu\nu}R^{(1)}=-{\kappa\over 2}(T^{mat}_{\mu\nu}+T^{grav}_{\mu\nu})\label{eq:fc}
\end{equation}
where we identify
\begin{equation}
T^{grav}_{\mu\nu}={4\over \kappa^2}(R^{(2)}_{\mu\nu}-{1\over 2}\eta_{\mu\nu}R^{(2)})-{2\over \kappa} h_{\mu\nu}R^{(1)}
\end{equation}
as the energy-momentum tensor of the gravitational field.  Using
\begin{eqnarray}
R_{\mu\nu}^{(2)}&=&\kappa^2\left\{
-{1\over 4}\partial_\mu h_{\alpha\beta}\partial_\nu
h^{\alpha\beta} -{1\over 2}\partial_\alpha
h_{\mu\lambda}\partial^\alpha h^\lambda_\nu+{1\over 2}\partial_\alpha h_{\mu\lambda}
\partial^\lambda h^{\alpha}_\nu\right.\nonumber\\
&+&\left.{1\over 2}h^{\lambda\alpha}\left[\partial_\lambda\partial_\nu
h_{\mu\alpha}+\partial_\lambda\partial_\mu h_{\nu\alpha}-
\partial_\mu\partial_\nu h_{\lambda\alpha}-\partial_\lambda
\partial_\alpha h_{\mu\nu}\right]\right.\nonumber\\
&+&\left.{1\over 2}\left((\partial\cdot h)^\alpha-{1\over 2}
\partial^\alpha h\right)\left(
\partial_\mu h_{\nu\alpha}+\partial_\nu h_{\mu\alpha}
-\partial_\alpha h_{\mu\nu}\right)\right\}\nonumber\\
R^{(2)}&=&\eta^{\mu\nu}R^{(2)}_{\mu\nu}=\kappa^2\left\{-{3\over 4}\partial_\mu
h_{\alpha\beta}\partial^\mu h^{\alpha\beta}+{1\over
2}\partial_\alpha h_{\mu\lambda}
\partial^\lambda h^{\mu\alpha}\right.\nonumber\\
&+&\left.{1\over 2}h^{\lambda\alpha}\left(2\partial_\lambda\left((\partial\cdot h)_\alpha-{1\over 2}\partial_\alpha h\right)
-\Box h_{\lambda\alpha}\right)
\right.\nonumber\\
&+&\left.\left((\partial\cdot h)^\alpha-{1\over 2}
\partial^\alpha h\right)
\left((\partial\cdot h)_\alpha-{1\over 2}
\partial_\alpha h\right)\right\}
\end{eqnarray}
we find
\begin{eqnarray}
T_{\mu\nu}^{\rm grav}&=&-2h^{\lambda\kappa}\left(
\partial_\mu\partial_\nu
h_{\lambda\kappa}+\partial_\lambda\partial_\kappa h_{\mu\nu}
-\partial_\kappa\left(\partial_\nu
h_{\mu\lambda}+\partial_\mu
h_{\nu\lambda}\right)\right)\nonumber\\
&-&2\partial_\lambda h_{\sigma\nu}\partial^\lambda
h^{\sigma}_\mu +2\partial_\lambda
h_{\sigma\nu}\partial^\sigma h^{\lambda}_\mu -\partial_\nu
h_{\sigma\lambda}\partial_\mu
h^{\sigma\lambda}\nonumber\\
&-&\eta_{\mu\nu}\left(\partial_\lambda h_{\sigma\chi}
\partial^\sigma h^{\lambda\chi}
-{3\over 2}\partial_\lambda h_{\sigma\chi}\partial^\lambda
h^{\sigma\chi}-h^{\alpha\beta}\Box h_{\alpha\beta}\right)
-h_{\mu\nu}\Box h\nonumber\\
\quad\label{eq:gn}
\end{eqnarray}

Before proceeding we also need to deal with the gauge-invariance by using the Faddeev-Popov
method\cite{fpm,thv}, leading to a second order action of the form

\begin{eqnarray}
S^{(2)}_{tot}&=&\int d^4x \left[-{3\over 2}\partial_\mu
h_{\alpha\lambda}\partial^\mu h^{\alpha\lambda}+\partial_\alpha h_{\mu\lambda}
\partial^\lambda h^{\mu\alpha}\right.\nonumber\\
&-&\left.h^{\lambda\alpha}\Box h_{\lambda\alpha}
-\left(\partial^\beta h^\alpha_\beta-{1\over 2}
\partial^\alpha h\right)
\left(\partial^\sigma h_{\sigma\alpha}-{1\over 2}
\partial_\alpha h\right)\right]\nonumber\\
&+&S_{ghost}(\eta_\mu)
\end{eqnarray}
where $\eta_\mu$ is a fermion ghost field.

Using the harmonic gauge condition Eq. (\ref{eq:tr}) and liberally integrating by parts we find
\begin{eqnarray}
S^{(2)}_{tot}&=&{1\over 4}\int d^4x h^{\mu\nu}(x)\left(\eta_{\mu\alpha}\eta_{\nu\beta}+\eta_{\mu\beta}\eta_{\nu\alpha}-\eta_{\mu\nu}\eta_{\alpha\beta}
\right)\Box h^{\alpha\beta}(x)+S_{ghost}(\eta_\mu)\nonumber\\
&=&{1\over 2}\int d^4x h^{\mu\nu}(x)P_{\mu\nu,\alpha\beta}\Box h^{\alpha\beta}(x)+S_{ghost}(\eta_\mu)
\end{eqnarray}
where
\begin{equation}
P_{\mu\nu,\alpha\beta}={1\over 2}(\eta_{\mu\alpha}\eta_{\nu\beta}+\eta_{\mu\beta}\eta_{\nu\alpha}-\eta_{\mu\nu}\eta_{\alpha\beta})
\end{equation}
Inverting, we see that the harmonic gauge graviton propagator is given by
\begin{equation}
G_{\mu\nu,\alpha\beta}(q)={iP_{\mu\nu,\alpha\beta}\over q^2+i\epsilon}
\end{equation}

With the propagator in hand we can begin to explore quantum gravitational effects provided we know the interaction.  In the electromagnetic case we have
the equation
\begin{equation}
\Box A_\mu+\partial_\mu(\partial_\nu A^\nu)=-eJ_\mu
\end{equation}
which, using the Lorentz gauge condition---$\partial_\nu A^\nu=0$---corresponds to the interaction Lagrangian
\begin{equation}
{\cal L}_{int}^{em}(x)=-eJ_\mu(x)A^\mu(x)
\end{equation}
Likewise in the case of the gravitational interaction we have the field equation Eq. (\ref{eq:jq})
which, using the harmonic gauge condition---$\partial^\beta h_{\beta\mu}-{1\over 2}\partial_\mu h=0$---corresponds to the interaction Lagrangian
\begin{equation}
{\cal L}_{int}^{grav}(x)={\kappa\over 2}T_{\mu\nu}(x)h^{\mu\nu}(x)
\end{equation}
so that the gravitational "charge" and "current" are $\kappa/2$ and $T_{\mu\nu}$ respectively. It is clear then that in order to determine the graviton couplings, we need to know the matrix elements of the energy-momentum tensor.

We begin by considering the case of a scalar field, for which the energy-momentum tensor is derived in Appendix A---the lowest order scalar energy-momentum vertex is given by
\begin{equation}
<p_2|T_{\mu\nu}^{(0)}(x)|p_1>={e^{i(p_2-p_1)\cdot x}\over \sqrt{4E_1E_2}}\left[2P_\mu P_\nu-{1\over 2}(q_\mu q_\nu-\eta_{\mu\nu}q^2)\right]\label{eq:cx}
\end{equation}
where $P={1\over 2}(p_1+p_2)$ is the mean energy-momentum and $q=p_1-p_2$ is the momentum transfer.  Note that the energy-momentum tensor is conserved in that
\begin{equation}
q^\mu<p_2|T_{\mu\nu}(x)|p_1>=0
\end{equation}
as required by taking the divergence of the field equation Eq. (\ref{eq:fx}) and using the harmonic gauge condition.
Of course, radiative corrections---gravitational or electromagnetic---lead to modifications of the lowest order matrix element Eq. (\ref{eq:cx}), which must have the general form
\begin{equation}
<p_2|T_{\mu\nu}(x)|p_1>={e^{i(p_2-p_1)\cdot x}\over \sqrt{4E_1E_2}}\left[2P_\mu P_\nu F_1(q^2)+(q_\mu q_\nu-\eta_{\mu\nu}q^2)F_2(q^2)\right]\label{eq:em}
\end{equation}
As can be seen from the condition
\begin{equation}
<p_2|\hat{P}_\mu|p_1>=P_\mu<p_2|p_1>=<p_2|\int d^3x T_{\mu 0}(x)|p_1>=P_\mu F_1(q^2=0)<p_2|p_1>
\end{equation}
conservation of energy-momentum requires
\begin{equation}
F_1(q^2=0)=1
\end{equation}
but there exists no constraint on $F_2(q^2)$.  We see from Eq. (\ref{eq:cx}) that at lowest order $F_1=1$ and $F_2=-{1\over 2}$, so that this condition is satisfied.

\section{Photonic loops for the Reissner-Nordstrom and Kerr-Newman metrics}

Before examining the modifications induced by gravitational corrections, it is useful to examine a related but simpler case---electromagnetic radiative corrections to $T_{\mu\nu}$---in order to understand the relevant physics\cite{hbg,ecl}.  Calculationally the photon loop corrections are very similar to those arising from graviton loops because photons and gravitons are both massless and propagate long distances. The example of the photon will introduce the nonanalytic corrections that occur in momentum space and will show how these are correspondingly nonlocal in position space.

Our subject is the photon loop correction to the energy-momentum tensor of a charged particle, which will reveal the form of the gravitational field in the vicinity. The classical result should be the Reissner-Nordstrom metric and Kerr-Newman metric, which describe the gravitational field around a charged particle without and with spin.  Because of the charge, the photon loop is the leading correction. Gravity can be treated classically here.

In treating the photon loop, we need the gravitational coupling to the photon. In lowest order this involves
the electromagnetic energy-momentum tensor
\begin{equation}
^{cl}T_{\mu\nu}^{em}={2\over \sqrt{-g}}{\partial\over \partial g_{\mu\nu}}(\sqrt{-g}{\cal L}_{em})=-F_{\mu\beta}{F_\nu}^\beta+{1\over 4}\eta_{\mu\nu}F_{\beta\gamma}F^{\beta\gamma}\label{eq:hg}
\end{equation}
which leads to a lowest order energy-momentum tensor for the photon
\begin{eqnarray}
\left\langle p_{2},\epsilon_{2}\right|T_{\mu\nu}(x)\left|p_{1},\epsilon_{1}\right\rangle  & = & \frac{e^{i(p_{2}-p_{1})\cdot x}}{\sqrt{4E_{2}E_{1}}}\left[2P_{\mu}P_{\nu}\epsilon_{1}\cdot\epsilon_{2}^*\right.\nonumber\\
 & + & P_{\mu}\left(\epsilon_{2\nu}^*\epsilon_{1}\cdot q-\epsilon_{1\nu}\epsilon_{2}^*\cdot q\right)+P_{\nu}\left(\epsilon_{2\mu}^*\epsilon_{1}\cdot q-\epsilon_{1\mu}\epsilon_{2}^*\cdot q\right)\nonumber\\
 & - & \left.\frac{1}{2}\left(q_{\mu}q_{\nu}-\eta_{\mu\nu}q^{2}\right)\epsilon_{1}\cdot\epsilon_{2}^*-\eta_{\mu\nu}\epsilon_{1}\cdot q\epsilon_{2}^*\cdot q\right.\nonumber\\
 & + & \left.\frac{q_{\mu}}{2}\epsilon_{2\nu}^*\epsilon_{1}\cdot q+\frac{q_{\nu}}{2}\epsilon_{2\mu}^*\epsilon_{1}\cdot q+\frac{q_{\mu}}{2}\epsilon_{1\nu}\epsilon_{2}^*\cdot q+\frac{q_{\nu}}{2}\epsilon_{1\mu}\epsilon_{2}^*\cdot q\right.\nonumber\\
 & - & \left.\frac{q^{2}}{2}\left(\epsilon_{1\mu}\epsilon_{2\nu}^*+\epsilon_{2\mu}^*\epsilon_{1\nu}\right)\right]\label{eq:dg}
\end{eqnarray}
Eq. (\ref{eq:dg}) defines the Feynman rule for the photon coupling.

We can now evaluate the various photon loop diagrams, which will lead to modifications of the lowest order spin zero form factors. The scalar electromagnetic vertices are well known and, using the photon-photon-graviton vertex given in Eq. (\ref{eq:dg}), the results are\cite{hbg}
\begin{eqnarray}
^0F_1^{em}(q^2)&=&1+{\alpha_{em}\over 4\pi}{q^2\over m^2}\left(-{8\over 3}+{3\over 4}{\pi^2 m\over \sqrt{-q^2}}+2\log{-q^2\over m^2}-{2\over 3}\log{\lambda^2\over m^2}\right)+\ldots\nonumber\\
^0F_2^{em}(q^2)&=&-{1\over 2}+{\alpha_{em}\over 4\pi}\left(-{2\over \epsilon}+\gamma+\log{m^2\over 4\pi\mu^2}-{26\over 9}+{1\over 2}{\pi^2 m\over \sqrt{-q^2}}+{4\over 3}\log{-q^2\over m^2}\right)+\ldots\nonumber\\
\quad
\end{eqnarray}
where $\alpha_{em}=e^2/4\pi$ is the fine structure constant.  Here we have used dimensional regularization and $\lambda$ is a photon ``mass'' which is inserted in order to regulate the infrared sector of the theory and which disappears when bremsstrahlung corrections are included.  We observe that there exist two types of radiative corrections here.  One class is analytic and therefore local when the transition to coordinate space is made.  (This includes the ultraviolet divergence, which can be absorbed into the coefficient of a term $RF_{\mu\nu}F^{\mu\nu}{\rm Tr} QUQU^\dagger$ in the effective Lagrangian.)  The second class is more interesting and involves nonanalytic terms such as $\sqrt{-q^2}$ and $q^2\log{-q^2}$.  Such forms do not occur in radiative corrections to the electromagnetic current and arise here from the triangle and bubble diagrams in Figures 1a and 1b involving coupling of the energy-momentum tensor to a pair of photons.  It is the presence of the {\it two} massless propagators in such diagrams, and the fact that both photons can be nearly on shell, which leads to this nonanalytic structure\cite{jbh}.

\begin{figure}[ht]
\centerline{
\includegraphics[width=.5\textwidth]{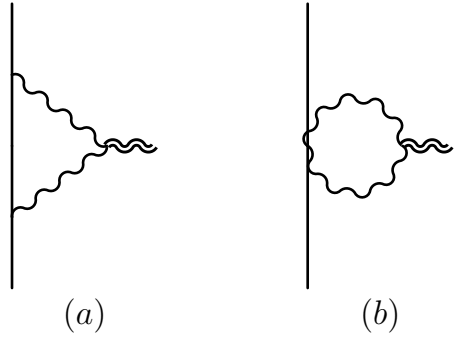}
}
\caption{Electromagnetic correction diagrams having nonanalytic components. Here
the single wiggly lines represent photons while the double
wiggly line indicates coupling to a graviton.. }
\label{EM}
\end{figure}

The physics of these nonanalytic terms can be easily extracted by making the transition to coordinate space in the Breit frame, wherein $q_0=0$ and $\boldsymbol{p}_1=-\boldsymbol{p}_2=\boldsymbol{q}/2$---
\begin{eqnarray}
^0T_{00}^{em}(\boldsymbol{r})&=&\int{d^3q\over (2\pi)^3}e^{i\boldsymbol{q}\cdot\boldsymbol{r}}\left(m ^0F_1(-\boldsymbol{q}^2)+{\boldsymbol{q}^2\over 2m} ^0F_2(-\boldsymbol{q}^2)\right)\nonumber\\
&=&\int{d^3q\over (2\pi)^3}e^{i\boldsymbol{q}\cdot\boldsymbol{r}}\left[m-{\pi\alpha_{em }|\boldsymbol{q}|\over 8}-{\alpha_{em}\boldsymbol{q}^2\over 3\pi m}\log{\boldsymbol{q}^2\over m^2}\right]+\ldots\nonumber\\
&=&m\delta^3(\boldsymbol{r})+{\alpha_{em}\over 8\pi r^4}-{\alpha_{em}\hbar\over \pi^2 mr^5}+\ldots\nonumber\\
^0T_{i0}^{em}(\boldsymbol{r})&=& 0\nonumber\\
^0T_{ij}^{em}(\boldsymbol{r})&=&{1\over 2m}\int{d^3q\over (2\pi)^3}e^{i\boldsymbol{q}\cdot\boldsymbol{r}}(q_iq_j-\delta_{ij}\boldsymbol{q}^2)^0F_2(-\boldsymbol{q}^2)\nonumber\\
&=&\int{d^3q\over (2\pi)^3}e^{i\boldsymbol{q}\cdot\boldsymbol{r}}(q_iq_j-\delta_{ij}\boldsymbol{q}^2)\left({\pi\alpha_{em}\over 16|\boldsymbol{q}|}+{\alpha_{em}\over 3\pi m}\log{\boldsymbol{q}^2\over m^2}\right)+\ldots\nonumber\\
&=&-{\alpha_{em}\over 4\pi r^4}\left({r_ir_j\over r^2}-{1\over 2}\delta_{ij}\right)-{\alpha_{em}\hbar\over 3\pi^2mr^5}+\ldots\nonumber\\
\quad\label{eq:xz}
\end{eqnarray}
where we have restored the factor of $\hbar$ in order to differentiate classical and quantum mechanical contributions and the ellipses denote short distance pieces.  The interesting feature here is that a quantum loop diagram has generated a classical effect---that is, a term independent of $\hbar$.  As mentioned above this is due to the presence of two massless propagators in the diagrams\cite{jbh}.  The meaning of these classical pieces is quite clear and can be understood by using the classical electromagnetic energy-momentum tensor which results from the lowest order
electromagnetic Lagrangian.  In the vicinity of a stationary particle having charge e and located at the origin we have $\boldsymbol{E}(\boldsymbol{r})=e\hat{r}/4\pi r^2$ so that
\begin{eqnarray}
^{cl}T_{00}^{em}(\boldsymbol{r})&=&{1\over 2}\boldsymbol{E}^2={\alpha_{em}\over 8\pi r^4}\nonumber\\
^{cl}T_{i0}^{em}(\boldsymbol{r})&=&0\nonumber\\
^{cl}T_{ij}^{em}(\boldsymbol{r})&=&-E_iE_j+{1\over 2}\delta_{ij}\boldsymbol{E}^2=-{\alpha\over 4\pi r^4}\left({r_ir_j\over r^2}-{1\over 2}\delta_{ij}\right)
\end{eqnarray}
which agree precisely with the classical components of Eq. (\ref{eq:xz}).  The meaning of the $\hbar$-dependent corrections can be understood qualitatively by including the effects of zitterbewegung.  The point is that at a classical level the distance of the probe from the particle can taken to be a fixed distance $r$.  However, including quantum mechanical effects, the location of the source is uncertain by an amount of order the Compton wavelength---$\delta r\sim \hbar/m<<r$.  Thus $1/r^4$ should be replaced by the form $1/(r+\delta r)^4\sim {1\over r^4}-{4\hbar\over mr^5}$ which has the form of the quantum corrections found above.

The feature that the loop correction leads to well-understood classical corrections is also valid if the source particle has spin.  In the case of spin 1/2 the form of the radiative corrections to the energy-momentum tensor have also been calculated.  In this case, as shown in Appendix A, the lowest order energy-momentum tensor vertex has the form
\begin{equation}
<p_2|T_{\mu\nu}^{(0)}(x)|p_1>=e^{i(p_2-p_1)\cdot x}{1\over 4}\bar{u}(p_2)\left[\gamma_\mu P_\nu+\gamma_\nu P_\mu-\eta_{\mu\nu}(\not\!{P}-m)\right]u(p_1)
\end{equation}
After radiative corrections the energy-momentum tensor can be written in the general form
\begin{eqnarray}
<p_2|T_{\mu\nu}(x)|p_1>&=&e^{i(p_2-p_1)\cdot x}\bar{u}(p_2)\left[{1\over m}P_\mu P_\nu\, ^{1\over 2}F_1(q^2)+{1\over m}(q_\mu q_\nu-\eta_{\mu\nu}q^2)^{1\over 2}F_2(q^2)\right.\nonumber\\
&-&\left.\left({i\over 4m}\sigma_{\mu\beta}q^\beta P_\nu+{i\over 4m}\sigma_{\nu\beta}q^\beta P_\mu\right) ^{1\over 2}F_3(q^2)\right]u(p_1)
\end{eqnarray}
so we see that in lowest order we have
\begin{equation}
^{1\over 2}F_1^{(0)}= ^{1\over 2}F_3^{(0)}=1\qquad ^{1\over 2}F_2^{(0)}=0\label{eq:nc}
\end{equation}
In the spin 1/2 case, besides the constraint of energy-momentum conservation discussed above we have the additional requirement of angular momentum conservation.  Defining
\begin{equation}
\hat{M}_{12}=\int d^3x(T_{01}x_2-T_{02}x_1)\stackrel{q\rightarrow 0}{\longrightarrow}-i\nabla_{q_2}\int d^3xe^{i\boldsymbol{q}\cdot\boldsymbol{r}}T_{01}(\boldsymbol{r})+i\nabla_{q_1}\int d^3xe^{i\boldsymbol{q}\cdot\boldsymbol{r}}T_{02}(\boldsymbol{r}),
\end{equation}
since
\begin{equation}
{1\over 2}=\lim_{q\rightarrow 0}<p_2,\uparrow|\hat{M}_{12}|p_1,\uparrow>=\bar{u}_\uparrow(p){1\over 2}\sigma_3u_\uparrow(p)^{1\over 2}F_3(q^2=0)
\end{equation}
we see that angular momentum conservation requires that $^{1\over 2}F_3(q^2=0)=1$.  Again energy-momentum conservation requires $^{1\over 2}F_1(q^2=0)=1$ while there is no constraint on $^{1\over 2}F_2(q^2)$.  Obviously the lowest order forms given in Eq. (\ref{eq:nc}) satisfy these conditions.

Performing now the loop integrations associated with electromagnetic corrections, we determine that
\begin{equation}
^{1\over 2}F_1^{em}(q^2)=^0F_1^{em}(q^2)\quad{\rm and}\quad ^{1\over 2}F_2^{em}(q^2)=^0F_2^{em}(q^2)
\end{equation}
while for the new form factor $^{1\over 2}F_3^{em}(q^2)$
\begin{equation}
^{1\over 2}F_3^{em}(q^2)=1+{\alpha_{em}\over 4\pi}{q^2\over m^2}\left(-{47\over 18}+{1\over 2}{m\pi^2\over \sqrt{-q^2}}+{2\over 3}\log{-q^2\over m^2}-{2\over 3}\log{\lambda^2\over m^2}\right)
\end{equation}
For the spin 1/2 energy-momentum tensor we find
\begin{eqnarray}
^{1\over 2}T_{00}^{em}(\boldsymbol{r})&=&\int{d^3q\over (2\pi)^3}e^{i\boldsymbol{q}\cdot\boldsymbol{r}}\left(m ^{1\over 2}F_1(-\boldsymbol{q}^2)+{\boldsymbol{q}^2\over 2m}
^{1\over 2}F_2(-\boldsymbol{q}^2)\right)\nonumber\\
&=&m\delta^3(\boldsymbol{r})+{\alpha_{em}\over 8\pi r^4}-{\alpha_{em}\hbar\over \pi^2 mr^5}+\ldots\nonumber\\
^{1\over 2}T_{i0}^{em}(\boldsymbol{r})&=&i\int{d^3q\over (2\pi)^3}e^{i\boldsymbol{q}\cdot\boldsymbol{r}}{1\over 2}\boldsymbol{S}\times\boldsymbol{q}_i ^{1\over 2}F_3(-\boldsymbol{q}^2)\nonumber\\
&=&i\int{d^3q\over (2\pi)^3}e^{i\boldsymbol{q}\cdot\boldsymbol{r}}{1\over 2}\boldsymbol{S}\times\boldsymbol{q}_i\left(1-{\alpha_{em}\pi\over 8m}|\boldsymbol{q}|-{\alpha_{em}\boldsymbol{q}^2\over 6\pi m^2}\log{\boldsymbol{q}^2\over m^2}\right)\nonumber\\
&=&{1\over 2}(\boldsymbol{S}\times\boldsymbol{\nabla})_i\delta^3(\boldsymbol{r})-\left({\alpha_{em}\over 4\pi mr^6}-{5\alpha_{em}\hbar\over 4\pi^2m^2r^7}\right)(\boldsymbol{S}\times\boldsymbol{r})_i\nonumber\\
^{1\over 2}T_{ij}^{em}(\boldsymbol{r})&=&{1\over 2m}\int{d^3q\over (2\pi)^3}e^{i\boldsymbol{q}\cdot\boldsymbol{r}}(q_iq_j-\delta_{ij}\boldsymbol{q}^2)\, ^{1\over 2}F_2(-\boldsymbol{q}^2)\nonumber\\
&=&-{\alpha_{em}\over 4\pi r^4}\left({r_ir_j\over r^2}-{1\over 2}\delta_{ij}\right)-{\alpha_{em}\hbar\over 3\pi^2mr^5}+\ldots
\label{eq:vz}
\end{eqnarray}
where $\boldsymbol{S}={1\over 2}\chi_f^\dagger\boldsymbol{\sigma}\chi_i$.  That is, the diagonal spin zero and spin 1/2 energy-momentum tensor densities are identical---
$$^{1\over 2}T_{00}^{em}(\boldsymbol{r})=^0T_{00}^{em}(\boldsymbol{r})\quad{\rm and}\quad ^{1\over 2}T_{ij}^{em}(\boldsymbol{r})=^0T_{ij}^{em}(\boldsymbol{r})$$
---but there now exists a nonzero off-diagonal term $^{1\over 2}T_{i0}^{em}(\boldsymbol{r})$.  The form of the classical correction in this term can be understood from the feature that a Dirac particle has both an electric field {\it and} magnetic field
\begin{equation}
\boldsymbol{E}={e\hat{r}\over 4\pi r^2}\quad{\rm and}\quad \boldsymbol{B}={e\over m}{3\hat{r}\boldsymbol{S}\cdot\hat{r}-\boldsymbol{S}\over 4\pi r^3}
\end{equation}
The off-diagonal piece of the classical energy-momentum tensor density, Eq. (\ref{eq:hg})
\begin{equation}
^{cl}T_{i0}=-(\boldsymbol{E}\times\boldsymbol{B})_i
\end{equation}
then becomes
\begin{equation}
^{cl}T_{i0}(\boldsymbol{r})=-{\alpha_{em}\over 4\pi mr^6}(\boldsymbol{S}\times \boldsymbol{r})_i
\end{equation}
in agreement with the classical component found in Eq. (\ref{eq:vz}).  The form of the corrected energy-momentum tensor has also been calculated for a spin 1 particle, yielding identical results as for the spin 1/2 case, except for the replacements $\chi_f^\dagger\chi\rightarrow\hat{\epsilon}_f^*\cdot\hat{\epsilon}_i$ and ${1\over 2}\chi_f^\dagger\boldsymbol{\sigma}\chi_i\rightarrow i\hat{\epsilon}_f\times\hat{\epsilon}_i$ plus new quadrupole corrections\cite{spo}.  We suspect that the forms of these corrections---both classical and quantum---are universal.

Before moving the gravitational case it is useful to note one other interesting feature in the electromagnetic case.  Using the linearized Einstein equation---Eq. \ref{eq:td}---and the results for the energy-momentum tensor $T_{\mu\nu}(\boldsymbol{r})$ generated above we can solve for the metric tensor, which yields the form
\begin{eqnarray}
^0h_{00}(\boldsymbol{r})&=&-{2Gm\over r}+{G\alpha_{em}\over r^2}-{8G\alpha_{em}\hbar\over 3\pi mr^3}+\ldots\nonumber\\
^0h_{i0}(\boldsymbol{r})&=&0\nonumber\\
^0h_{ij}(\boldsymbol{r})&=&-\delta_{ij}{2Gm\over r}+{G\alpha_{em}r_ir_j\over r^4}+{4G\alpha_{em}\hbar\over 3\pi mr^3}\left({r_ir_j\over r^2}-\delta_{ij}\right)+\ldots\label{eq:fd}
\end{eqnarray}
for a spinless particle and
\begin{eqnarray}
^{1\over 2}h_{00}(\boldsymbol{r})&=&-{2Gm\over r}+{G\alpha_{em}\over r^2}-{8G\alpha_{em}\hbar\over 3\pi mr^3}+\ldots\nonumber\\
^{1\over 2}h_{i0}(\boldsymbol{r})&=&\left({2G\over r^3}-{G\alpha_{em}\over mr^4}+{2G\alpha_{em}\hbar\over \pi m^2r^5}\right)(\boldsymbol{S}\times\boldsymbol{r})_i+\ldots\nonumber\\
^{1\over 2}h_{ij}(\boldsymbol{r})&=&-\delta_{ij}{2Gm\over r}+{G\alpha_{em}r_ir_j\over r^4}+{4G\alpha_{em}\hbar\over 3\pi mr^3}\left({r_ir_j\over r^2}-\delta_{ij}\right)+\ldots\label{eq:sa}
\end{eqnarray}
for a particle with spin 1/2.  (The results for spin 1 have also been calculated and agree with those of spin 1/2 up to small quadrupole corrections\cite{spo} so again, it is likely these results too are universal\cite{spo}.)  The classical components of the spin zero results Eq. (\ref{eq:fd}) agree precisely with those of the Reissner-Nordstrom metric\cite{rnm}, which is the metric associated with a massive charged  particle, while the spin 1/2 results Eq. (\ref{eq:sa}) agree with those of the Kerr-Newman metric\cite{knm}, which is the metric associated with a massive charged particle which is spinning.  Again the form of the quantum mechanical corrections are consistent with zitterbewegung fluctuations.

\section{Gravitational correction to the Schwarzschild metric}

We now repeat the same procedure but with gravitational loops. This amounts to looking at graviton loop corrections to the Schwarzschild metric. This procedure is not gauge invariant, and even the concept of a metric is not a fully quantum concept. However, the result is an illustration of the form of quantum corrections. We are working in harmonic gauge and the result applies only in that gauge. However, in the process of calculating the quantum correction, we also obtain the first classical correction to Schwarzschild, a result first found by Duff\cite{duf}.

Again there exist nonanalytic forms arising from the triangle and bubble diagrams containing two massless propagators---Figure 2---which lead to classical and quantum mechanical corrections to the lowest order results.  Using the gravitational couplings given in Appendix A and keeping only the the nonanalytic pieces, the gravitationally corrected form factors in the case of a spinless particle of mass $m$ are found to be
\begin{eqnarray}
^0F_1^{grav}(q^2)&=&1+{Gq^2\over \pi}\left({1\over 16}{\pi^2 m\over \sqrt{-q^2}}-{3\over 4}\log{-q^2\over m^2}\right)+\ldots\nonumber\\
^0F_2^{grav}(q^2)&=&-{1\over 2}+{Gm^2\over \pi}\left({7\over 8}{\pi^2 m\over \sqrt{-q^2}}-2\log{-q^2\over m^2}\right)+\ldots .
\end{eqnarray}
The corresponding energy-momentum tensor is
\begin{eqnarray}
^0T_{00}^{grav}(\boldsymbol{r})&=&\int{d^3q\over (2\pi)^3}e^{i\boldsymbol{q}\cdot\boldsymbol{r}}\left(m ^0F_1^{grav}(-\boldsymbol{q}^2)+{\boldsymbol{q}^2\over 2m} ^0F_2^{grav}(-\boldsymbol{q}^2)\right)\nonumber\\
&=&m\delta^3(\boldsymbol{r})-{3Gm^2\over 8\pi r^4}-{3Gm\hbar\over 4\pi^2r^5}\nonumber\\
^0T_{i0}^{grav}(\boldsymbol{r})&=&0\nonumber\\
^0T_{ij}^{grav}(\boldsymbol{r})&=&{1\over 2m}\int{d^3q\over (2\pi)^3}e^{i\boldsymbol{q}\cdot\boldsymbol{r}}(q_iq_j-\delta_{ij}\boldsymbol{q}^2)^0F_2^{grav}(-\boldsymbol{q}^2)\nonumber\\
&=&-{7Gm^2\over 4\pi r^4}\left({r_ir_j\over r^2}-{1\over 2}\delta_{ij}\right)+{2Gm\hbar\over \pi^2r^5}\delta_{ij}\label{eq:hj}
\end{eqnarray}
In this case we can compare the classical results with the predictions of the gravitational energy-momentum tensor density---Eq. (\ref{eq:gn})---
\begin{eqnarray}
^{cl}T_{00}^{grav}(\boldsymbol{r})&=&\left(-3\boldsymbol{\nabla}f(r)\cdot\boldsymbol{\nabla}f(r)-12f(r)\boldsymbol{\nabla}^2f(r)\right)=-{3Gm^2\over 8\pi r^4}+\ldots\nonumber\\
^{cl}T_{i0}^{grav}(\boldsymbol{r})&=&0\nonumber\\
^{cl}T_{ij}^{grav}(\boldsymbol{r})&=&\left(-2\nabla_if(r)\nabla_jf(r)+3\delta_{ij}\boldsymbol{\nabla}f(r)\cdot\boldsymbol{\nabla}f(r)
-4f(r)\nabla_i\nabla_jf(r)\right.\nonumber\\
&+&\left.4\delta_{ij}f(r)\boldsymbol{\nabla}^2f(r)\right)=-{7Gm^2\over 4\pi r^4}\left({r_ir_j\over r^2}-{1\over 2}\delta_{ij}\right)+\ldots
\end{eqnarray}
where the ellipses denote short distance components, which agree precisely with the classical component of the loop calculation result---Eq. (\ref{eq:hj}).


\begin{figure}[ht]
\begin{center}
\epsfig{file=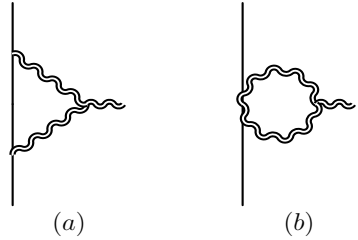,height=4cm,width=7cm} \caption{Feynman
diagrams having nonanalytic components.  Here
 the doubly wiggly lines represent gravitons.}
\end{center}
\end{figure}

\medskip

In the case of spin 1/2, the gravitationally corrected form factors are found to be
\begin{eqnarray}
^{1\over 2}F_1^{grav}(q^2)&=&1+{Gq^2\over \pi}\left({1\over 16}{\pi^2 m\over \sqrt{-q^2}}-{3\over 4}\log{-q^2\over m^2}\right)+\ldots\nonumber\\
^{1\over 2}F_2^{grav}(q^2)&=&-{1\over 2}+{Gm^2\over \pi}\left({7\over 8}{\pi^2 m\over \sqrt{-q^2}}-2\log{-q^2\over m^2}\right)+\ldots\nonumber\\
^{1\over 2}F_3^{grav}(q^2)&=&1+{Gq^2\over \pi}\left({1\over 4}{\pi^2 m\over \sqrt{-q^2}}+{1\over 4}\log{-q^2\over m^2}\right)+\ldots
\end{eqnarray}
That is, the spin zero and spin 1/2 values for the form factors $F_1,F_2$ are identical---
$$^{1\over 2}F_1^{grav}(q^2)=^0F_1^{grav}(q^2)\quad {\rm and} \quad ^{1\over 2}F_2^{grav}(q^2)=^0F_2^{grav}(q^2)$$
---which in turn implies that the diagonal components of the energy-momentum tensor densities are are identical to those of spin zero---
$$^0T_{00}(\boldsymbol{r})=^{1\over 2}T_{00}(\boldsymbol{r})\quad {\rm and}\quad ^0T_{ij}(\boldsymbol{r})=^{1\over 2}T_{ij}(\boldsymbol{r})$$
However, there is now a nonzero off-diagonal piece
\begin{equation}
^{1\over 2}T_{i0}^{grav}(\boldsymbol{r})={1\over 2}(\boldsymbol{S}\times\boldsymbol{\nabla})_i\delta^3(\boldsymbol{r})-\left({Gm\over 2\pi r^6}-{15G\hbar\over 4\pi^2r^7}\right)(\boldsymbol{S}\times\boldsymbol{r})_i
\end{equation}
where $\boldsymbol{S}={1\over 2}\chi_f^\dagger\boldsymbol{\sigma}\chi_i$, which can be compared to the predictions from the classical gravitational energy-momentum tensor density---Eq. (\ref{eq:gn})---
\begin{eqnarray}
^{cl}T_{i0}^{grav}(\boldsymbol{r})&=&{2\over m}\left[-(\boldsymbol{S}\times\boldsymbol{\nabla})_jf(r)\nabla_i\nabla_jf(r)+\nabla_jf(r)\nabla_i(\boldsymbol{S}\times\boldsymbol{\nabla})_jf(r)\right]\nonumber\\
&=&-{Gm\over 2\pi r^6}(\boldsymbol{S}\times\boldsymbol{r})_i
\end{eqnarray}
Again there is complete agreement between the two classical calculations.  (The spin 1 calculation has also been performed and agrees with the spin 1/2 forms, except for small quadrupole corrections\cite{spo}, so that we suspect that these results are universal.)

As in the electromagnetic case, we can use the linearized Einstein equation to convert the results for the energy-momentum tensor to those for the metric tensor, yielding for spin zero
\begin{eqnarray}
^0h_{00}^{grav}(\boldsymbol{r})&=&-{2Gm\over r}+{2G^2m^2\over r^2}+{7G^2m\hbar\over \pi r^3}+\ldots\nonumber\\
^0h_{i0}^{grav}(\boldsymbol{r})&=&0\nonumber\\
^0h_{ij}^{grav}(\boldsymbol{r})&=&-\delta_{ij}{2GM\over r}-{G^2m^2\over r^2}\left({r_ir_j\over r^2}+\delta_{ij}\right)-{G^2m\hbar\over \pi r^3}\left(8{r_ir_j\over r^2}+\delta_{ij}\right)+\ldots\nonumber\\
\quad
\end{eqnarray}
the classical components of which agree completely with the Schwarzschild metric\cite{scm}, which characterizes a stationary massive particle.

In the case of spin 1/2 the diagonal components of the metric tensor are identical to the spin zero case---$^{1\over 2}h_{00}^{grav}(\boldsymbol{r})=^0h_{00}^{grav}(\boldsymbol{r})$ and
$^{1\over 2}h_{ij}^{grav}(\boldsymbol{r})=^0h_{ij}^{grav}(\boldsymbol{r})$---but there exists now a nonzero off-diagonal component.
\begin{equation}
^{1\over 2}h_{i0}^{grav}(\boldsymbol{r})=\left({2G\over r^3}-{2G^2m\over r^4}+{3G^2\hbar\over \pi r^5}\right)(\boldsymbol{S}\times\boldsymbol{r})_i
\end{equation}
The classical components of the spin 1/2 metric tensor agree completely with the Kerr metric\cite{kem}, which characterizes a massive spinning particle.  Again, the spin 1 result has also been calculated and agrees completely with the spin 1/2 forms\cite{spo}, so we suspect that they are universal.

\section{Correction to the Newton Potential}

It has long been thought to be an unattainable goal to calculate effect of quantum physics on the gravitational interaction. However, for the long distance quantum correction to the Newtonian potential, the result is both simple and universal. We explain the logic in this section.

There are several possible definitions of a gravitational potential. We shall discuss some of the associated subtleties below, but for the moment we shall simply define the potential $V(r)$ as the Fourier transform of the nonrelativistic scattering amplitude ${\cal M}(\boldsymbol{q})$---
\begin{equation}
V(r)=-\int{d^3q\over (2\pi)^3}e^{i\boldsymbol{q}\cdot\boldsymbol{r}}{\cal M}(\boldsymbol{q})
\end{equation}
and we begin our discussion by considering the scattering of a pair of spinless particles.

\subsection{Spin 0-Spin 0 Scattering}

At lowest order the interaction of two spinless particles of mass $m_1,\,m_2$ is described in terms of the one graviton exchange (tree) amplitude
\begin{eqnarray}
^0{\cal M}^{(1)}(q)&=&-i{1\over \sqrt{2E_12E_22E_32E_4}}{1\over 2}\kappa \left[p_{1\alpha}p_{3\beta}+p_{1\beta}p_{3\alpha}-\eta_{\alpha\beta}(p_3\cdot p_1-m_1^2)\right]{iP^{\alpha\beta,\gamma\delta}\over q^2}\nonumber\\
&\times&{1\over 2}\kappa \left[p_{2\gamma}p_{4\delta}+p_{2\delta}p_{4\gamma}-\eta_{\gamma\delta}(p_4\cdot p_2-m_2^2)\right]\nonumber\\
&=&{-8\pi G\over \sqrt{2E_12E_22E_32E_4}}\left[{(s-m_1^2-m_2^2+{1\over 2}q^2)^2-2m_1^2m_2^2-{1\over 4}q^4\over q^2}\right]
\end{eqnarray}
We shall utilize the symmetric center of mass
frame with incoming momenta $\boldsymbol p_1 = -\boldsymbol{p}_2=\boldsymbol p + \boldsymbol q
/ 2$ and outgoing momenta $\boldsymbol p_3 = -\boldsymbol{p}_4=\boldsymbol p - \boldsymbol q / 2$. Conservation of energy then requires $\boldsymbol p
\cdot \boldsymbol q = 0$ so that $\boldsymbol p_i^{\hspace*{1.4pt} 2} = \boldsymbol
p^{\hspace*{1.4pt} 2} + \boldsymbol q^{\hspace*{1.4pt} 2} / 4$ for $i = 1,
2, 3, 4$ and $q^2 = - \boldsymbol q^{\hspace*{1.4pt} 2}$. In the
nonrelativistic limit--- $\boldsymbol q^{\hspace*{1.4pt} 2}, \boldsymbol
p^{\hspace*{1.4pt} 2} \ll m^2$ ---the lowest order amplitude reads
\begin{eqnarray}
{}^0 \! {\cal M}^{(1)}(\boldsymbol q) & \simeq & \frac{4 \pi G m_1 m_2} {\boldsymbol q^{\hspace*{1.4pt} 2}}
\left[1 + \frac{\boldsymbol p^{\hspace*{1.4pt} 2}}{m_1 m_2} \left(1+\frac{3(m_1 + m_2)^2}{2 m_1 m_2}\right)
+ \ldots\right] \nonumber \\
&+& G \pi \left[\frac{3(m_1^2 + m_2^2)}{2 m_1 m_2} + \frac{\boldsymbol p^{\hspace*{1.4pt} 2}}{m_1 m_2}
\left(3 - \frac{5(m_1^2 + m_2^2)^2}{4 m_1^2 m_1^2}\right) + \ldots \right] + \ldots \nonumber \\ \label{eq_ampLO_00}
\end{eqnarray}
yielding the potential
\begin{eqnarray}
{}^0V^{(1)}_G(\boldsymbol{r})&=&-\int {d^3q\over (2\pi)^3} \,
{}^0 \!{\cal M}^{(1)}(\boldsymbol q) \, e^{-i\boldsymbol{q}\cdot\boldsymbol{r}} \nonumber\\
&=&- \hspace*{1pt} \frac{G m_1 m_2}{r} \left[1 + \frac{\boldsymbol p^{\hspace*{1.4pt} 2}}{m_1 m_2}
\left(1+\frac{3(m_1 + m_2)^2}{2 m_1 m_2}\right) + \ldots\right] \nonumber \\
&& + \, G \pi \delta^3(\boldsymbol r) \left[\frac{3(m_1^2 + m_2^2)}{2 m_1 m_2} +
\frac{\boldsymbol p^{\hspace*{1.4pt} 2}}{m_1 m_2}\left(3 - \frac{5(m_1^2 + m_2^2)^2}{4 m_1^2 m_2^2}\right) + \ldots \right]\label{eq:jg}
\nonumber\\
\label{eq:po}
\end{eqnarray}
and we recognize the Newtonian potential as the dominant piece of Eq. (\ref{eq:jg}) (accompanied by a small kinematic correction) together with a short range modification.

\begin{figure}
\begin{center}
\epsfig{file=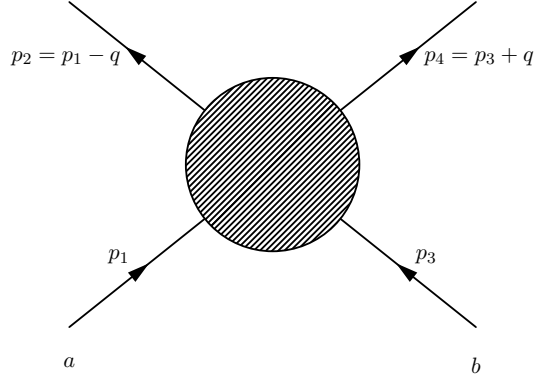,width=7cm} \caption{Basic kinematics of
gravitational scattering. } \label{fig_kinem}
\end{center}
\end{figure}

Our goal is to examine corrections to this lowest order potential due to two-graviton exchange and thereby to define a higher order gravitational potential.  This problem has been previously studied by Iwasaki using noncovariant perturbation theory\cite{iwa}, and by Khriplovich and Kirilin\cite{kka},\cite{kkb} and by Bjerrum-Bohr, Donoghue, and Holstein\cite{bdh} using conventional Feynman diagrams.  Our approach will be similar to that used in
\cite{kka},\cite{kkb} and \cite{bdh}.  The diagrams utilized are shown in Figure 3 and the various interaction vertices are derived in Appendix A so it is merely a matter of calculating these triangle and bubble diagrams.  Because of the many tensor indices involved, this is a challenging but straightforward problem and until recently there had been a number of mistakes in such evaluations\cite{mtk}, which have finally been corrected\cite{bdh}.  As before, the procedure is to calculate the various diagrams while retaining only the nonanalytic components since only these terms lead to long range corrections to the Newtonian potential.  The nonanalytic contributions which arise from the various diagrams can be expressed in terms of the quantities $L=\sqrt{\boldsymbol{q}}^2$ and $S=\pi^2/ \sqrt|\boldsymbol{q}|$
Summing all the scattering diagrams, we determine the total
\begin{equation}
{}^0{\cal M}^{(2)}_{tot}(q)=G^2m_1m_2\left[6(m_1+m_2)S - {41\over
5}L \right]-i4\pi G^2m_1^2m_2^2 \hspace*{1pt} {L\over
q^2}\sqrt{m_1m_2\over s-s_0}\label{eq:kp}
\end{equation}
and observe that, in addition to the expected terms involving $L$
and $S$, there arises a component of the second order amplitude which is
{\it imaginary} and represents a scattering phase.  The origin of this imaginary term is
from the second Born approximation to the Newtonian potential, and
suggests that in order to define a proper correction to the first
order Newtonian potential we must subtract off such pieces.  For this
purpose we work in the nonrelativistic limit and the center of
mass frame---$\boldsymbol{p}_1+\boldsymbol{p}_2=0$---as defined above.  We have then
\begin{equation}
s-s_0=2\sqrt{m_1^2+ \boldsymbol{p}_1^{\hspace*{1.4pt} 2}} \sqrt{m_2^2+\boldsymbol{p}_1^{\hspace*{1.4pt} 2}}
+2 \boldsymbol{p}_1^{\hspace*{1.4pt} 2} -2m_1m_2
\end{equation}
and
\begin{equation}
\sqrt{m_1m_2\over s-s_0}\simeq {m_r\over p_0}
\end{equation}
where $m_r=m_1m_2/(m_1+m_2)$ is the reduced mass and $p_0 \equiv
|\boldsymbol{p}_i|, \  i=1,2,3,4$.  The transition amplitude Eq.
(\ref{eq:kp}) then assumes the form
\begin{equation}
{}^0{\cal M}^{(2)}_{tot}(\boldsymbol q) \simeq G^2m_1m_2\left[6(m_1+m_2)S -{41\over 5}L\right]-i4\pi G^2m_1^2m_2^2 \hspace*{1pt} {L\over q^2}{m_r\over p_0} .
\end{equation}

We can evaluate the Born iteration directly by utilizing the simple Newtonian potential
\begin{equation}
{}^0V^{(1)}_G(\boldsymbol r)=-{Gm_1m_2\over r}\label{eq:co}
\end{equation}
which reproduces the long distance behavior of the lowest order
amplitude for spin-0 -- spin-0 gravitational scattering---Eq.
(\ref{eq:po})---in the nonrelativistic limit.  The corresponding
momentum space representation is
\begin{equation}
 {}^0V^{(1)}_G(\boldsymbol q) \equiv \left<\boldsymbol p_f \left| {}^0 \hat V^{(1)}_G \right|
 \boldsymbol p_i \right> = - \frac{4 \pi G m_1 m_2}{\boldsymbol q^{\hspace*{1.4pt}2}} =
 - \frac{4 \pi G m_1 m_2}{(\boldsymbol{p}_i - \boldsymbol{p}_f)^2} \label{eq:lp}
\end{equation}
and the second Born term becomes
\begin{eqnarray} \label{eq:it}
{}^0{\rm Amp}_G^{(2)}(\boldsymbol q)&=&- \int{d^3\ell\over
(2\pi)^3} \, \frac{\left<\boldsymbol p_f \left| {}^0 \hat V^{(1)}_G \right| \boldsymbol \ell \,
\right> \left<\boldsymbol \ell \left| {}^0 \hat V^{(1)}_G \right|
\boldsymbol p_i \right>}{E(p_0) - E(\ell) + i \epsilon}\nonumber\\
&=&-i \hspace*{0.4pt} 4\pi G^2 m_1^2 m_2^2 \hspace*{1pt} \frac{L}{q^2} \frac{m_r}{p_0}\label{eq:ib}
\end{eqnarray}
which precisely reproduces the imaginary component of ${}^0{\cal
M}_{tot}^{(2)}(\boldsymbol q)$, as expected.  In order to produce a properly
defined second order potential ${}^0V^{(2)}_G(\boldsymbol r)$ we must
subtract this second order Born term from the second order
transition amplitude, yielding a well-defined second order gravitational potential
\begin{eqnarray}
{}^0V_G^{(2)}(\boldsymbol r)&=&-\int{d^3q\over
(2\pi)^3}e^{-i\boldsymbol{q}\cdot\boldsymbol{r}}\left[{}^0{\cal
M}_{tot}^{(2)}(q)-{}^0{\rm Amp}_G^{(2)}(q)\right]\nonumber\\
&=&\int{d^3q\over (2\pi)^3}e^{-i\boldsymbol{q}\cdot\boldsymbol{r}}
\hspace*{1pt} G^2 m_1 m_2\left[-6S(m_1+m_2) + {41\over 5}L\right]\nonumber\\
&=&-{3G^2m_1m_2(m_1+m_2)\over r^2
}-{41G^2m_1m_2\hbar\over 10\pi r^3}\nonumber\\
\quad\label{eq:so}
\end{eqnarray}

The quantum mechanical---$\sim \hbar/mr^3$---component of the second
order potential given in Eq. (\ref{eq:so}) agrees with that
previously given by Bjerrum-Bohr, Donoghue, and Holstein\cite{bdh}
and by Kirilin and Khriplovich\cite{kka}.
However, the classical---$\sim 1/r^2$---contribution quoted by
Iwasaki
\begin{equation}
 {}^0V_{IW}^{(2)}(\boldsymbol r)= {G^2m_1m_2(m_1+m_2)\over 2 r^2}.
\end{equation}
differs in both sign and magnitude from that quoted above in Eq. (\ref{eq:so}) and by
Bjerrum-Bohr et al. in \cite{bdh}.  The resolution of this issue has been
given by Sucher, who pointed out that the form of the classical interaction depends
upon the precise definition of the first order potential used in the
iteration\cite{jsu}. Moreover, it depends and on whether one uses
relativistic forms of the leading order potentials and the
nonrelativistic propagator $G^{(0)}(\ell)$ in the iteration.  In modern terms, the
potential depends on how one performs the matching---{\it e.g.}, Iwasaki\cite{iwa} performs an off-shell matching while we match
on-shell.\footnote{Besides the dependence on the forms used in the
iteration, the classical piece also depends on the coordinates used.
The quantum piece however depends neither on the choice of
coordinates\cite{bdh} nor on the iteration forms\cite{hre}.}. Use
of the simple lowest order form Eq. (\ref{eq:lp}) within a
nonrelativistic iteration yields our result for the
amplitude given in Eq. (\ref{eq:ib}) and is sufficient to
remove the offending imaginary component of the scattering amplitude. In
Appendix B we derive an alternative form of the
$\mathcal O(G^2)$ classical potential which results from an
iteration that includes the leading relativistic corrections and
which reproduces the Iwasaki result\cite{iwa}.

Therefore, a unique definition of the second order potential potential does not exist.
However, ambiguities in the form of the second order classical
potential are not a concern, since the potential is {\it not}
an observable.  What {\it is} an observable is the on-shell
transition amplitude, which is uniquely defined in each case as
\begin{equation}
{}^0 \!{\cal M}_{tot}(\boldsymbol q)=-\int
d^3re^{i\boldsymbol{q}\cdot\boldsymbol{r}}\left[{}^0V_i^{(1)}(\boldsymbol
r)+{}^0V_i^{(2)}(\boldsymbol r)\right]+{}^0{\rm Amp}_i(\boldsymbol{q})
\end{equation}
where the index $i$ denotes differing possible definitions of the
potentials and the iteration. Thus we regard the potential as merely a
way to display the resulting scattering amplitude in coordinate space,
and we emphasize that the main results are the long distance components
of the scattering amplitude---${}^0 \!{\cal M}_{tot}(\boldsymbol q)$.  With these caveats in mind, the total potential
describing the gravitational scattering of spinless particles, at second order in $G$, can be written as
\begin{equation}
^0V_{tot}^{(2)}(r)=-{Gm_1m_2\over r}\left(1+3{G(m_1+m_2)\over r}+{41G\hbar\over 10\pi r^2}\right)
\end{equation}
and we observe that there exist long range contributions to the leading Newtonian potential, with a classical component falling as $1/r^2$ together with a quantum mechanical corrections dropping as $\hbar/mr^3$.

\subsection{Spin 0-Spin ${1\over 2}$ Scattering}

The calculation of the correction to the Newtonian potential can also be carried out straightforwardly in the case of a spin 1/2 particle having mass $m_2$ scattering from a spinless particle of mass $m_1$.  The tree level transition amplitude from one-graviton exchange is
\begin{eqnarray}
{}^{1\over 2}{\cal M}^{(1)}(q) \hspace*{-5pt} &= \hspace*{-5pt}&{- 16 \pi G m_1 m_2 \over
\sqrt{2E_12E_2E_3E_4}}\Bigg[-\frac{m_1 m_2}{q^2} \hspace*{1pt} \bar{u}(p_4)u(p_2) \nonumber \\
&& \hspace*{76pt} {}+ \frac{s - m_1^2 - m_2^2 + \frac{1}{2} q^2}{q^2} \, {1\over m_1}\bar{u}(p_4)\!\not\!{p}_1u(p_2)\Bigg] . \quad
\end{eqnarray}
Defining the spin vector as
\begin{equation}
S_2^\mu=-{1\over 2}\bar{u}(p_4)\gamma^\mu\gamma_5 u(p_2)\underset{NR}{\longrightarrow}\chi_{2f}^\dagger{1\over 2}\boldsymbol{\sigma}\chi_{2i}
\end{equation}
the nonanalytic part of the transition amplitude in the threshold
limit $s \rightarrow s_0 = (m_1 + m_2)^2$ can be written in the form
\begin{equation}
{}^{1\over 2}{\cal M}^{(1)}(q) \simeq -{4\pi G m_1 m_2\over
q^2}\left[\bar{u}(p_4)u(p_2)+{2 i\over
m_1m_2^2}\epsilon_{\alpha\beta\gamma\delta}p_1^\alpha p_2^\beta
q^\gamma S_2^\delta \right].
\end{equation}
In order to define the potential we again take the nonrelativistic
amplitude in the symmetric center of mass frame ($\boldsymbol p_1 = - \boldsymbol
p_2 = \boldsymbol p + \boldsymbol q /2$)---
\begin{equation}
{}^{1\over 2}{\cal M}^{(1)}(\boldsymbol q)\simeq {4\pi Gm_1m_2\over
\boldsymbol{q}^{\hspace*{1.4pt} 2}}\left[\chi^\dagger_{2f}\chi_{2i} + {i(3m_1 + 4 m_2)\over
2m_1m_2^2}\boldsymbol{S}_2\cdot\boldsymbol{p}\times\boldsymbol{q}+\ldots\right]
\end{equation}
and the lowest order potential becomes
\begin{eqnarray}
{}^{1\over 2}V_G^{(1)}(\boldsymbol r)&=&-\int {d^3q\over (2\pi)^3} \, \hspace*{1pt}
{}^{1\over 2}{\cal M}^{(1)}(\boldsymbol q \hspace*{1pt}) \, e^{-i\boldsymbol{q}\cdot\boldsymbol{r}}\nonumber\\
&=&-{G m_1 m_2 \over r}\chi^\dagger_{2f}\chi_{2i}
-{3 m_1 + 4 m_2\over
2 m_1 m_2^2}\boldsymbol{S}_2\cdot\boldsymbol{p}\times\boldsymbol{\nabla}\left(-{G m_1 m_2 \over r}\right)\nonumber\\
&=&-{Gm_1m_2\over r}\chi^\dagger_{2f}\chi_{2i}+{G\over
r^3}{3m_1 + 4m_2\over 2m_2}\boldsymbol{L}\cdot\boldsymbol{S}_2\label{eq:oj}
\end{eqnarray}
where $\boldsymbol{L}=\boldsymbol{r}\times\boldsymbol{p}$ is the angular momentum---the
modification of the leading spin-independent potential has a
spin-orbit character.

A subtlety that
arises in the calculation involving spin is that {\it two}
independent kinematic variables arise: the momentum transfer $q^2$
and $s - s_0$, which is to leading order proportional to $p_0^2$
(where $p_0^2 \equiv \boldsymbol p_i^{\hspace*{1.4pt}2},  \ i=1,2,3,4$) in
the center of mass frame.  We find that our results differ if we
perform an expansion first in $s - s_0$ and then in $q^2$ or vice
versa. This ordering issue occurs only for the box diagram, diagram
(d) of Fig. 2, where it stems from the reduction of
vector and tensor box integrals.  Their reduction in terms of scalar
integrals involves the inversion of a matrix whose Gram determinant
vanishes in the nonrelativistic threshold limit $q^2, s - s_0
\rightarrow 0$. More precisely, the denominators or the vector and
tensor box integrals (see Appendix A in \cite{hre}) involve a
factor of $(4 p_0^2 - \boldsymbol q^{\hspace*{1.4pt} 2})$ when expanded in
the nonrelativistic limit. Since $q^{\hspace*{1.4pt} 2} = 4 p_0^2
\sin^2 \frac{\theta}{2}$ with $\theta$ the scattering angle, we
notice that $4 p_0^2 > \boldsymbol q^{\hspace*{1.4pt} 2}$ unless we
consider backward scattering where $\theta = \pi$ and where the
scattering amplitude diverges. And since $p_0^2$ originates from the
relativistic structure $s - s_0$, it is clear that one must first
expand our vector and tensor box integrals in $q^2$ and then in $s -
s_0$.

Calculating the various diagrams as before we find the total second order contribution
\begin{eqnarray}
{}^{1\over 2}{\cal M}_{tot}^{(2)}(q) \hspace*{-5pt} &=\hspace*{-5pt}&G^2 m_1 m_2 \Bigg[
\bar{u}(p_4)u(p_2) \bigg(6(m_1 + m_2) S - \frac{41}{5} L\bigg) \nonumber \\
&& \hspace*{41pt} + \hspace*{0.5pt} \frac{i}{m_1 m_2^2} \hspace*{1pt} \epsilon_{\alpha\beta\gamma\delta} \hspace*{1pt} p_1^\alpha p_2^\beta q^\gamma S_2^\delta \,
 \bigg(\frac{11(3 m_1 + 4 m_2)}{4} S - \frac{64}{5} L\bigg) \nonumber\\
&& \hspace*{41pt}+ \hspace*{0.5pt} {i S (3 m_1 + 4 m_2)\over
m_2 (s-s_0)}\epsilon_{\alpha\beta\gamma\delta}
p_1^\alpha p_2^\beta q^\gamma S_2^\delta\Bigg]\nonumber\\
&&-i4\pi G^2 m_1^2 m_2^2 \hspace*{1pt} {L\over q^2}\sqrt{m_1 m_2\over
s-s_0}\left(\bar{u}(p_4)u(p_2)+ {2 i\over
m_1m_2^2}\epsilon_{\alpha\beta\gamma\delta}
p_1^\alpha p_2^\beta q^\gamma S_2^\delta\right) \nonumber\\
\quad \label{eq:ohla}
\end{eqnarray}
Finally, working in the center of mass frame and taking the
nonrelativistic limit, Eq. (\ref{eq:ohla}) becomes
\begin{eqnarray}
{}^{1\over 2}{\cal M}_{tot}^{(2)}(\boldsymbol q)\hspace*{-5pt} &\simeq\hspace*{-5pt}&\left[G^2 m_1 m_2
\left( 6 (m_1 + m_2) S - {41 \over 5}L\right) - i 4 \pi G^2 m_1^2 m_2^2 \hspace*{1pt} \frac{L}{q^2} \frac{m_r}{p_0}\right]
\chi^\dagger_{2f}\chi_{2i} \nonumber\\
&+\hspace*{-5pt}&\Bigg[G^2  \hspace*{-1pt} \left(\frac{12m_1^3 \hspace*{-1pt} + \hspace*{-1pt} 45 m_1^2 m_2 \hspace*{-1pt} + \hspace*{-1pt} 56 m_1 m_2^2
 \hspace*{-1pt} + \hspace*{-1pt} 24 m_2^3}{2(m_1 \hspace*{-1pt} + \hspace*{-1pt} m_2)}  \hspace*{1pt} S
- \frac{87 m_1 \hspace*{-1pt} + \hspace*{-1pt} 128 m_2}{10}  \hspace*{1pt} L\right)\nonumber\\
&&+ \frac{G^2 m_1^2 m_2^2 (3 m_1 + 4 m_2)}{(m_1 + m_2)} \left(- i \frac{2 \pi
L}{p_0 q^2} + \frac{S}{p_0^2} \right)\Bigg] {i\over
m_2}\boldsymbol{S}_2\cdot\boldsymbol{p}\times\boldsymbol{q}\label{eq:yo}
\end{eqnarray}
We observe from Eq. (\ref{eq:yo}) that the scattering amplitude
consists of two pieces---
\begin{itemize}
\item[i)] a spin-independent component proportional
to $\chi^\dagger_{2f}\chi_{2i}$ whose functional form
\begin{equation}
 G^2 m_1 m_2 \left( 6 (m_1 + m_2) S - {41 \over 5}L\right) - i 4 \pi G^2 m_1^2 m_2^2 \hspace*{1pt} \frac{L}{q^2} \frac{m_r}{p_0}
\end{equation}
is {\it identical} to that of spinless scattering,
\item[ii)] a spin-orbit component proportional to
$${i\over m_2}\boldsymbol{S}_2\cdot\boldsymbol{p}\times\boldsymbol{q}$$
whose functional form is
\begin{eqnarray}
&&G^2 \left(\frac{12m_1^3 + 45 m_1^2 m_2 + 56 m_1 m_2^2 + 24 m_2^3}{2(m_1 + m_2)}  \hspace*{1pt} S
- \frac{87 m_1 + 128 m_2}{10}  \hspace*{1pt} L\right)\nonumber\\
& + & \frac{G^2 m_1^2 m_2^2 (3 m_1 + 4 m_2)}{(m_1 + m_2)} \left(- i \frac{2 \pi
L}{p_0 q^2} + \frac{S}{p_0^2} \right) \label{eq:im}
\end{eqnarray}
\end{itemize}
We note in Eq. (\ref{eq:im}) the presence in the spin-orbit
potential of an imaginary final state rescattering term proportional
to $i/p_0$, similar to that found in the case of spin-independent
scattering, together with a {\it completely new} type of kinematic
form, proportional to $1/p_0^2$ which diverges at threshold.  The
presence of {\it either} term would prevent us from writing down a
well defined second order potential.

The solution to this problem is, as before, to properly subtract the
iterated first order potential---
\begin{eqnarray}
{}^{1\over 2}{\rm Amp}_G^{(2)}(\boldsymbol q) &=&-\sum_{n} \int{d^3\ell\over
(2\pi)^3} \, \frac{\left<\boldsymbol p_f,\chi_f \left| {}^{\frac{1}{2}} \hat
V^{(1)}_G \right| \boldsymbol {\ell},\chi_{n} \right> \left<\boldsymbol{\ell},\chi_{n} \left|
{}^{\frac{1}{2}} \hat V^{(1)}_G \right| \boldsymbol{p}_i,\chi_i
\right>}{{p_0^2\over 2m_r} - \frac{\ell^2}{2 m_r} + i \epsilon}\nonumber\\
\quad
\end{eqnarray}
where
\begin{equation}
\left<\boldsymbol{p}_f,\chi_f \left| {}^{\frac{1}{2}} \hat V^{(1)}_G \right| \boldsymbol
{p}_i,\chi_i \right> =
 \left<\boldsymbol p_f,\chi_f \left| {}^{\frac{1}{2}} \hat V^{(1)}_{S-I} \right| \boldsymbol p_i,\chi_i \right>
 + \left<\boldsymbol p_f,\chi_f \left| {}^{\frac{1}{2}} \hat V^{(1)}_{S-O} \right| \boldsymbol p_i,\chi_i \right>
\end{equation}
with
\begin{eqnarray}
 \left<\boldsymbol p_f,\chi_f \left| {}^{\frac{1}{2}} \hat V^{(1)}_{S-I} \right| \boldsymbol p_i,\chi_i \right>
 &=& - {4 \pi G m_1 m_2 \over \boldsymbol{q}^{\hspace*{1.4pt}2}} \, \chi^\dagger_{2f}\chi_{2i}
  = - \frac{4 \pi G m_1 m_2}{(\boldsymbol p_i - \boldsymbol p_f)^2} \, \chi^\dagger_{2f}\chi_{2i} \nonumber\\
 \left<\boldsymbol p_f,\chi_f \left| {}^{\frac{1}{2}} \hat V^{(1)}_{S-O} \right| \boldsymbol p_i,\chi_i \right>
 &=&- {4 \pi G m_1 m_2 \over \boldsymbol{q}^{\hspace*{1.4pt}2}}{3 m_1+ 4 m_2\over 2m_1m_2}\, \frac{i}{m_2}\boldsymbol{S}_2\cdot\boldsymbol{p}\times\boldsymbol{q} \nonumber\\
 &=&- {4 \pi G m_1 m_2 \over (\boldsymbol p_i - \boldsymbol p_f)^2}{3 m_1 + 4 m_2\over 2m_1m_2}\, \frac{i}{m_2}\boldsymbol{S}_2\cdot \frac{1}{2} (\boldsymbol p_i + \boldsymbol p_f) \times (\boldsymbol p_i - \boldsymbol p_f) \nonumber\\
\end{eqnarray}
We find that the iterated amplitude splits into
spin-independent and spin-dependent pieces.  The leading
spin-independent amplitude is
\begin{eqnarray}
{}^{1\over 2}{\rm Amp}^{(2)}_{S-I}(\boldsymbol q) \hspace*{-3pt} &= \hspace*{-3pt} &-\sum_{n}
\int{d^3\ell\over (2\pi)^3} \, \frac{\left<\boldsymbol p_f,\chi_f \left|
{}^{\frac{1}{2}} \hat V^{(1)}_{S-I} \right| \boldsymbol \ell,\chi_{n} \right>
\left<\boldsymbol \ell,\chi_{n} \left| {}^{\frac{1}{2}}
 \hat V^{(1)}_{S-I} \right| \boldsymbol p_i,\chi_i \right>}{\frac{p_0^2}{2 m_r} -
 \frac{\ell^2}{2 m_r} + i \epsilon} \nonumber\\
&=&-i4\pi G^2 m_1^2
m_2^2 \hspace*{1pt} {L\over
q^2}\frac{m_r}{p_0}\chi^\dagger_{2f}\chi_{2i}
\label{eq:iteration0hSI}
\end{eqnarray}
and the leading spin-dependent term is
\begin{eqnarray}
{}^{1\over 2}{\rm Amp}^{(2)}_{S-O}(\boldsymbol q) \hspace*{-6pt} &= \hspace*{-6pt}&-\sum_{n}
\int{d^3\ell\over (2\pi)^3} \, \frac{\left<\boldsymbol p_f,\chi_f \left|
{}^{\frac{1}{2}} \hat V^{(1)}_{S-I}
 \right| \boldsymbol{\ell},\chi_{n} \, \right> \left<\boldsymbol{\ell},\chi_{n} \left| {}^{\frac{1}{2}}
 \hat V^{(1)}_{S-O} \right| \boldsymbol p_i,\chi_i \right>}{\frac{p_0^2}{2 m_r} -
 \frac{\ell^2}{2 m_r} + i \epsilon} \nonumber\\
& = \hspace*{-6pt} &\frac{G^2 m_1^2 m_2^2 (3 m_1 + 4 m_2)}{(m_1 + m_2)} \left(- i \frac{2 \pi
L}{p_0 q^2} + \frac{S}{p_0^2} \right) {i\over
m_2}\boldsymbol{S}_2\cdot\boldsymbol{p}\times\boldsymbol{q} \label{eq:er}
\end{eqnarray}
so that when the amplitudes Eqs.
(\ref{eq:er}) and (\ref{eq:iteration0hSI}) are subtracted
from the full one loop scattering amplitude Eq. (\ref{eq:yo}) both
the terms involving $1/p_0^2$ and those proportional to $i/p_0$
disappear, leaving behind a well-defined second order potential
\begin{eqnarray}
{}^{1\over 2}V^{(2)}_{tot}(\boldsymbol r)&=&-\int{d^3q\over
(2\pi)^3}e^{-i\boldsymbol{q}\cdot\boldsymbol{r}}\left[{}^{1\over 2}{\cal
M}_{tot}^{(2)}(\boldsymbol{q})-{}^{1\over 2}{\rm
Amp}_G^{(2)}(\boldsymbol{q})\right]\nonumber\\
&=& \int{d^3q\over
(2\pi)^3}e^{-i\boldsymbol{q}\cdot\boldsymbol{r}}
\Bigg[G^2 m_1 m_2 \left(- 6(m_1+m_2)S+{41\over 5}L\right)
\chi_{2f}^\dagger\chi_{2i}\nonumber\\
&=&\left[-{3G^2m_1m_2(m_1+m_2)\over r^2} - {41G^2m_1m_2\hbar\over 10\pi r^3}\right]
\chi^\dagger_{2f}\chi_{2i}\nonumber\\
&+&\Bigg[{G^2(12 m_1^3 + 45 m_1^2 m_2 + 56 m_1 m_2^2 + 24 m_2^3)\over
2 m_2 (m_1+m_2)r^4} \nonumber \\
&&+{3 G^2(87 m_1+ 128m_2)\hbar\over 20 \pi
m_2 r^5}\Bigg] \boldsymbol{L}\cdot\boldsymbol{S}_2 \label{eq:mn}
\end{eqnarray}
We observe that the second order potential for long range
gravitational scattering of a spinless and spin-1/2 particle
consists of two components:  one independent of the spin of
particle 2 and identical to the potential found for the case of
spinless scattering, accompanied by a spin-orbit interaction
involving a new shorter range form for its classical and quantum
components.

\subsection{Universality}

We have seen that the leading quantum correction to the Newtonian potential is independent of the type of particle being considered. This result obtains despite the fact that very different Feynman diagrams occur when dealing with fermions and bosons. In fact, one can prove that that this universality is itself a low energy theorem of quantum gravity\cite{bxa}.

The argument which demonstrates that this equality is more than an accident has been shown by the use of some of the new methods of quantum field theory, which also serve as a check on the Feynman diagram calculation. The new techniques are often referred to as unitarity methods\cite{dxn}, because they rely on the unitarity and analyticity properties of Feynman diagrams. They work by identifying the unitarity cut in the amplitude and reconstructing the full Feynman diagram from this information. All one loop Feynman diagrams can be reduced to scalar bubble, triangle and box diagrams without any factors in the numerator of the loop integral, a property referred to as Passarino-Veltman reduction. Each of these structures has distinctive unitarity cuts. From the cuts then, one can reconstruct the prefactors of the box, triangle and bubble diagrams. In addition there can be polynomial terms which do not lead to cuts but, as we have above, we are interested only in the nonanalytic terms, which can be reconstructed properly.

For the calculations described above, we need only take the gravitational Compton amplitude---the coupling of two on-shell gravitons to two on-shell matter particle---and multiply them together in order to get the two-graviton unitarity cut. The contraction is performed most simply using helicity methods\cite{hlt}, which involve a form of axial gauge.

A second modern miracle further simplifies the amplitude method. It has recently been discovered that on-shell gravity amplitudes are in a precisely specified way related to on-shell gauge theory amplitudes\cite{fct}. This property is summarized by the phrase: gravity is the square of a gauge theory. In our cases, the gravitational Compton amplitude is the square, with a given prefactor, of the QED Compton amplitudes. The gravitational amplitude is very complex because of the presence of the triple graviton vertex, but the QED analog can be worked out straightforwardly by any field theory student.

These calculations have been carried out\cite{bxa}, reproducing the results of the Feynman diagram approach. This is gratifying as it shows both that the calculations have been done correctly and also that they are gauge invariant, as the two methods use different gauges. However, they also provide a proof in the universality low energy theorem. This is because the Compton amplitudes are already known to have universal soft limits\cite{wbf}. In the product, the leading term is then universal - this is the one that gives the classical correction. The quantum result follows from a term in the product which is a factor of $\sqrt{-q^2}$ higher than the leading term. However, the non-universal features appear only at a power $q^2$ higher than the leading term. This implies that the leading quantum correction is also universal.

It is worth mentioning that both the classical and quantum corrections have also been reproduced by dispersion relations methods, as reported in \cite{bxa}. This is another technique which reconstructs the real parts of the amplitudes from their on-shell cuts. The non-analytic terms that we are interested in are independent of the number of subtractions needed. The dispersive method has been done in both the harmonic gauge, in which case one needs to include cuts from ghost fields, and in the axial gauge of the helicity method, which has no ghosts. Again, this calculation can be used to prove the universality low energy theorem.

\section{Gravitational Scattering of a Massless System from a Massive System}

The calculations described above have dealt with the influence of gravity on massive systems and the results were based on an expansion
in powers of momentum transfer over mass.  However, gravity also couples to massless systems such as the photon. Here the calculations are
more complicated because both the photon and the graviton can propagate long distances in loops. The first published calculations
on these systems were described in \cite{bhr} and used the unitarity based methods described in the previous section. However, there
is also an unpublished thesis \cite{blk} which had studied the same systems\footnote{A few mistakes in the thesis were corrected by \cite{bhr}.} using
conventional Feynman diagrams. We will use the notation from both descriptions in what follows.

\subsection{Interactions of a massless scalar}

For simplicity we
begin with the case of a (fictitious) massless scalar field.

Consider a massless scalar particle of energy $E$ which is moving along the $z$-axis.  The lowest-order energy-momentum tensor of a massless scalar particle is still given by Eq. (\ref{eq:cx}), but now with $E_{1}=|\boldsymbol{p_{1}}|$,
$E_{2}= |\boldsymbol{p_{2}}|$. For the particle to act
like a fixed source, $p_{1}\approx p_{2}\approx p>>q$, and so
\begin{equation}
\left\langle p_{2}\right|T^{\mu\nu}\left(x\right)\left|p_{1}\right\rangle \approx E\hat{P}^{\mu}\hat{P}^{\nu}e^{i(p_2-p_1)\cdot x}\label{eq:ct}
\end{equation}
where $E\approx E_{1}\approx E_{2}$ is the time component of $P={1\over 2}(p_1+p_2)$
and $\hat{P}^\mu\equiv\frac{P^\mu}{E}$. Further,
\begin{eqnarray}
E_{1} & = & |\boldsymbol{p_{1}}| =|\boldsymbol{p}+\boldsymbol{q}/2| =\sqrt{\boldsymbol{p}^{2}+\boldsymbol{p}\cdot\boldsymbol{q}+\frac{\boldsymbol{q}^{2}}{4}}\approx|\boldsymbol{p}| +\frac{\boldsymbol{p}\cdot\boldsymbol{q}}{2|\boldsymbol{p}|}\nonumber \\
E_{2} & = & |\boldsymbol{p_{2}}| =|\boldsymbol{p}-\boldsymbol{q}/2| =\sqrt{\boldsymbol{p}^{2}-\boldsymbol{p}\cdot\boldsymbol{q}+\frac{\boldsymbol{q}^{2}}{4}}\approx|\boldsymbol{p}| -\frac{\boldsymbol{p}\cdot\boldsymbol{q}}{2|\boldsymbol{p}|}
\end{eqnarray}
so
\begin{equation}
q_{t}=E_{1}-E_{2}\approx\frac{\boldsymbol{p}}{|\boldsymbol{p}|}\cdot\boldsymbol{q}=q_{z},
\end{equation}
and $q\cdot x\approx q_{z}\left(t-z\right)-q_{x}x-q_{y}y$, where
the $z$-axis is taken to be in the direction of $\boldsymbol{P}$. Then,
transforming to coordinate space,
\begin{equation}
^0T^{\mu\nu}\left(x\right)=\int\frac{\mathrm{d}^{3}q}{\left(2\pi\right)^{3}} E\hat{P}^{\mu}\hat{P}^{\nu}e^{-i\left(q_{z}\left(z-t\right)+q_{x}x+q_{y}y\right)}
=E\delta\left(z-t\right)\delta\left(x\right)\delta\left(y\right)\hat{P}^{\mu}\hat{P}^{\nu}\label{eq:jw}
\end{equation}
which is the Aichelburg-Sexl form for the energy-momentum tensor
of a massless particle of energy $E$\cite{asx}.

\subsection{Metric Tensor}
The corresponding metric tensor is given by solving the (linearized) Einstein equation
\begin{equation}
\Box h_{\mu\nu}(q)=-16\pi G<p_2|T_{\mu\nu}(x)|p_1>
\end{equation}
where we have used the fact that ${\rm Tr} T=0$.   We have then
\begin{eqnarray}
h_{\mu\nu}(x)&=&-16\pi GE\hat{P}_\mu\hat{P}_\nu\int{d^3q\over (2\pi)^3}e^{i(q_z(z-t)+q_xx+q_yy)}{1\over q_x^2+q_y^2}\nonumber\\
&=&8GE\hat{P}_\mu\hat{P}_\nu\delta(z-t)\delta(\sqrt{x^2+y^2})\label{eq:as}
\end{eqnarray}
which is the form of the metric given by Aichelberg and Sexl\cite{asx}.

In order to calculate the gravitational loop corrections to Eq. (\ref{eq:jw}), we must evaluate the same diagrams as in the
massive case and the result can be written in terms of the two form factors $F_1(q^2),F_2(q^2)$ defined in Eq. (\ref{eq:em}), yielding
\begin{eqnarray}
F_1^{grav}(q^2)&=&1-{Gq^2\over 2\pi}\left(3\log({-q^2\over 4\pi\mu^2})+{1\over 2}\log^2({-q^2\over m^2})-\log({-q^2\over \lambda^2})\log({-q^2\over m^2})\right)\nonumber\\
F_2^{grav}(q^2)&=&{Gq^2\over 8\pi}\left(\log({-q^2\over 4\pi\mu^2})-{1\over 2}\log^2({-q^2\over m^2})+\log({-q^2\over \lambda^2})\log({-q^2\over m^2})\right)
\end{eqnarray}
where the particle mass $m^2$ has been used as a regulator.  Notice that there exist no classical nonanalyticities---$\sim \sqrt{-q^2}$---here.  This result is obvious in retrospect since there exists no mass scale to divide by and is consistent with the feature that Aichelberg and Sexl demonstrated explicitly that the linearized solution, Eq. (\ref{eq:as}), is also a solution of the full Einstein equation---there exist no higher order classical contributions\cite{asx}.

To lowest order in $q$ (and therefore at longest range) the energy-momentum tensor in co-ordinate space can naively be obtained by taking the Fourier transform, yielding
\begin{eqnarray}
T_{\mu\nu}^{grav}(x)&=&E\hat{P}_\mu\hat{P}_\nu\int{d^3q\over (2\pi)^3}e^{i(q_z(z-t)+q_xx+q_yy)}F_1^{grav}(q^2)+\ldots\nonumber\\
&=&E\hat{P}_\mu\hat{P}_\nu\delta(z-t)\left[\delta(x)\delta(y)+{8G\over (x^2+y^2)^2}\left(1+2\gamma+\log{\lambda^2(x^2+y^2)\over 4}\right)+\ldots\right]\nonumber\\
\quad\label{eq:hb}
\end{eqnarray}
However, there exists an obvious problem here in that Eq. (\ref{eq:hb}) depends on the artificial graviton mass $\lambda$.  Unlike the case of a massive particle, where the $\lambda$-dependence appears only in a short range term, the $\lambda$-dependence in Eq. (\ref{eq:hb}) is contained in a long range component and therefore we conclude that the long range energy-momentum tensor is not an observable.  This $\lambda$-dependence is {\it not} a problem, however, for the scattering amplitude, which {\it is} an observable and is the quantity which we next examine.

\subsection{Massless Particle Gravitational Scattering}

Consider now the gravitational scattering of a test particle of mass $m$ (later taken to be massless) from a heavy target mass $M$, both of which are taken to be spinless.  The needed diagrams are identical to those required for evaluation of the massive case, with the exception that we now must include the contributions from the bremsstrahlung terms---both from the massless ($m$) and massive ($M$) particles---which were unimportant in the massive scattering case since they were associated with short distance (analytic) effects.  However, there is an additional feature which must be addressed in the massless scattering situation, which
is the prevalence of infrared singularities. We know from QED that there are soft singularities which arise when loop momenta get small. Gravity has these
soft singularities also. In Yang-Mills theories with massless charged particle, there are also ``collinear divergences''\cite{ftb}, which lead to factors of $\log m^2$ and arise when one of the external momenta of the massless particles is parallel to a loop momentum.  In QED this collinear effect could in principle arise but, since there exist no massless charged particles, such $m\rightarrow 0$ divergences are not an issue.  In gravity, one might think that these singularities also exist as the massless gravitons carry their charge. However Weinberg
\cite{wbg} showed that gravity does not have collinear singularities, and so all that we need to deal with are the soft infrared divergences.

\subsection{Massless Particle-Massive Scalar Scattering: Result}

We begin with the gravitational interaction of a heavy scalar of mass $M$ with a light scalar of mass $m$, which will be later taken to vanish.  The elastic differential scattering cross section is then given by
\begin{equation}
\textrm{d}\sigma_{el}=\left(2\pi\right)^{4}\delta\left(p_{1}+p_{2}-p_{3}+p_{4}\right)\frac{\left|\mathcal{M}\right|^{2}}{4\sqrt{\left(p_{1}\cdot p_{2}\right)^{2}-M^{2}m^{2}}}\frac{\textrm{d}^{3}p_{3}\textrm{d}^{3}p_{4}}{\left(2\pi\right)^{3}2E_{3}\left(2\pi\right)^{3}2E_{4}}
\end{equation}
The calculation must be done very carefully and details are given in \cite{blk}.  However, the use of traditional Feynman diagram methods is quite tedious as well as challenging, and there exist a few errors in this result.  As an alternative then, Bjerrum-Bohr et al. performed the evaluation using modern on-shell helicity amplitude techniques\cite{bhr}, wherein the calculation is greatly simplified by at least three features:
\begin{itemize}
\item[a)] One is that the on-shell gravitational tree-level amplitudes can be written as the square of gauge theory amplitudes~\cite{kxq,hbh}.  In the case at hand the (nonabelian) gravitational Compton amplitudes are reduced to the product of (abelian) QED Compton amplitudes~\cite{bkj,bxa,hpv}.  The challenging diagrams involving the triple graviton vertex are avoided and are replaced by much simpler evaluations involving only QED vertices.  The general relation connecting the gravitational and electrodynamic Compton processes is derived in detail in ~\cite{hpv} and is given by
\begin{equation}\label{e:GravCompton}
 \hspace{-0.00cm} i {\cal M}{}^{[h(k_1)h(k_2)]}_{[\eta(p_1)\eta(p_2)]}= {\kappa^2\over 4e^2}\,{(p_1\cdot k_1)(p_1\cdot
    k_2)\over p_1\cdot p_2}\,  {\cal M}^{\rm QED}_{S=0}\, {\cal M}^{\rm QED}_{\eta}\!\!\!\!\,,
\end{equation}
where $h$ represents a graviton while $\eta$ can be either a photon $\gamma$ or massless spinless particle $\varphi$.  Here ${\cal M}^{\rm QED}_{\gamma}=\cM^{\rm QED}_{S=1}$ is the Compton amplitude for the scattering of a photon from a massless charged spin-1 target while $\cM^{\rm QED}_{\varphi}=\cM^{\rm QED}_{S=0}$ represents the Compton amplitude of a photon from a
massless charged spin-0 target.  These tree-level relations connect one-loop gravitational physics with one-loop electrodynamics in a non-trivial and interesting way~\cite{bxa}.

\item[b)] The second great simplification involves the use of on-shell unitarity techniques \cite{bcg}, instead of Feynman diagrams.  Unitarity-based calculations construct the relevant amplitude from the discontinuities of the scattering amplitude.  The long range nonanalytic terms in the one-loop amplitude can then be readily calculated from these on-shell cuts using the property of unitarity, as was directly demonstrated in ref.~\cite{bxa}. Cutting the graviton internal lines, the integrand of the one-loop amplitude factorizes in terms of a product of relatively simple tree amplitudes, given in this case by the gravitational Compton amplitudes.

\begin{figure}
\begin{center}
\epsfig{file=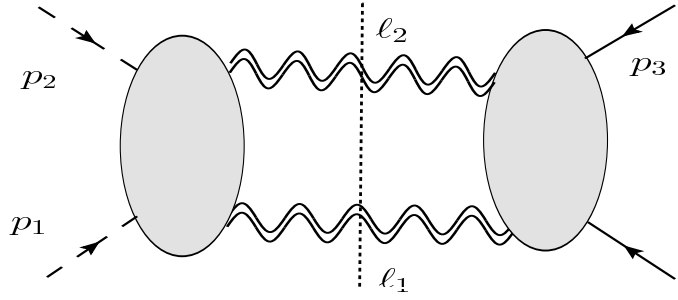,height=4cm,width=10cm}
\caption{\small The two graviton cut for the amplitude between a massless
   particle (dashed line) and a massive scalar (solid line). The
   grey blob are tree-level gravitational Compton amplitudes.}
\end{center}
\end{figure}

%
%
\item[c)] The final simplification is the use of the spinor-helicity formalism (see~\cite{mby} for a review). While this notation is perhaps less familiar, it drastically simplifies the form of the amplitudes which we display.
\end{itemize}

The tree-level massive scalar-graviton Compton amplitude is
\vskip-0.55cm
\begin{eqnarray}
  \label{e:4pointGrav}
 i {\cal M}{}^{[h^+(k_1)h^+(k_2)]}_{[\Phi(p_1)\Phi(p_2)]}&=&{\kappa^2\over 4}\,
 {M^4 \sq[k_1,k_2]^4\over (k_1\cdot k_2)( k_1\cdot p_1)(k_1\cdot p_2)} \,, \cr
 i \cMt{}^{[h^-(k_1)h^+(k_2)]}_{[\Phi(p_1)\Phi(p_2)]}&=&{\kappa^2\over 4}\,
{ \spab[k_1,p_1,k_2]^2\spab[k_1,p_2,k_2]^2\over (k_1\cdot k_2)( k_1\cdot p_1)(k_1\cdot p_2)}\,.
\end{eqnarray}
\vskip-0.1cm\noindent
The tree amplitudes connecting a massless scalar $\varphi$ and the graviton are then obtained by taking the limit $M\rightarrow 0$. Amplitudes with opposite helicity configurations are obtained by complex conjugation.  Computation of the cut discontinuity can be accomplished using traditional methods and is greatly simplified by the use of the on-shell identities.

Using these techniques the leading contribution to the amplitude (expanding all integrals in terms of leading order
contributions as done in~\cite{bkt,bxa}) is found to be:
\begin{eqnarray}\label{eq:gv}
&&    i
 {\cal M}^{[\phi(p_1)\phi(p_2)]}_{[\Phi(p_3)\Phi(p_4)]}
 \simeq (M\omega)^2\nonumber\\
&&\times
 \Big[{\kappa^2\over t}+\kappa^4 {15\over
  512}{ M\over \sqrt{-t}} \\ &&\nonumber
+\hbar \kappa^4
 {15\over
  512\pi^2}\,\log\left(-t\over M^2\right)
-\hbar\kappa^4\,{ bu^{\varphi} \over(8\pi)^2} \, \log\left(-t\over \mu^2\right)
\\
\nonumber &&+ \hbar\kappa^4 {3\over128\pi^2}\, \log^2\left(-t\over \mu^2 \right)
+\kappa^4 \,  {M\omega\over 8\pi} {i\over t}\log\left(-t\over
  M^2\right)
\Big]\,.
\end{eqnarray}
\vskip-0.1cm\noindent
where $\mu^2$ is the arbitrary mass scale parameter used in dimensional regularization and $bu^{\varphi}= 3/40$.

In Eq. \ref{eq:gv} the two terms in the second line correspond, respectively, to the leading Newtonian contribution and first post-Newtonian correction~\cite{ddn,bkt,bxa}.  The
next three (logarithmic) terms represent quantum gravity modifications.   The first term on the third line corresponds to
the quantum correction to the metric evaluated in~\cite{bks}.  The second piece on the third line
arises from the one-loop ultraviolet divergence of the amplitude and is the only contribution depending on the spin of the massless field.  On the fourth line the first term involves a new form not found in the previous (massive) analysis.  Finally, the last term, arising from the discontinuity of the box integral, contributes to the phase of the amplitude and is not directly observable. For this reason it will not be considered further.

One can straightforwardly generalize the calculation to the case that the massless scalars are replaced by photons.  In this case the only nonvanishing gravitational Compton helicity amplitudes involving photons $\gamma$ and gravitons $h$ are
\vskip-0.5cm
\begin{eqnarray}\label{e:ggGGhel}
\!\!\!\!\!\!\!\!\! i \overset{1}{\cal M}{}^{[h^+(k_1)h^-(k_2)]}_{[\gamma^+(p_1)\gamma^-(p_2)]}\!\!\!&=\!\!&{\kappa^2\over 4}{\sq[p_1,k_1]^2\an[p_2,k_2]^2\spab[k_2,p_1,k_1]^2\over
    (p_1\cdot p_2)(p_1\cdot k_1)(p_1\cdot k_2)},
\end{eqnarray}
\vskip-0.2cm\noindent
with $\overset{1}{\cal M}{}^{[h^+(k_1)h^-(k_2)]}_{[\gamma^-(p_1)\gamma^+(p_2)]}$ given by the above formula with $p_1$ and $p_2$ interchanged, and amplitudes with opposite helicity configurations are obtained by complex conjugation.  The resulting gravitational Compton amplitude involving a massive scalar and photon is then found to be

\begin{eqnarray}\label{eq:gw}
&&    i
 \cM^{[\gamma(p_1)\gamma(p_2)]}_{[\Phi(p_3)\Phi(p_4)]}
 \simeq \cN^{\gamma} (M\omega)^2\nonumber\\
&&\times
 \Big[{\kappa^2\over t}+\kappa^4 {15\over
  512}{ M\over \sqrt{-t}} \\ &&\nonumber
+\hbar \kappa^4
 {15\over
  512\pi^2}\,\log\left(-t\over M^2\right)
-\hbar\kappa^4\,{ bu^{\gamma} \over(8\pi)^2} \, \log\left(-t\over \mu^2\right)
\\
\nonumber &&+ \hbar\kappa^4 {3\over128\pi^2}\, \log^2\left(-t\over \mu^2 \right)
+\kappa^4 \,  {M\omega\over 8\pi} {i\over t}\log\left(-t\over
  M^2\right)
\Big]\,.
\end{eqnarray}
\vskip-0.1cm\noindent
where $bu^\gamma={-161/120}$ and $\cN^{\gamma}=(2M\omega)^{2} / (2\langle p_1|p_3|p_2]^2)$ for the $(+-)$ photon helicity contribution and its complex conjugate for the $(-+)$ photon helicity contribution.  The photon amplitude vanishes for the polarization configurations $(++)$ and $(--)$, which is a direct consequence of the properties of the tree-amplitudes in
eq.~\eqref{e:ggGGhel}.

In contrast with the non-relativistic case, where there was a universality theorem for the coefficient of the quantum correction, we see that most terms agree, except for one. Comparing Eqns. \ref{eq:gv} and \ref{eq:gw}, the exception is seen to be the $bu^{\eta} \,\log(-t/\mu^2)$ contribution from the massless bubble diagram ($\eta=\phi,\gamma$). 
We note also that, because of the vanishing of the photon scattering amplitudes for the helicity configurations $(++)$ and $(--)$, the amplitudes parallel and perpendicular to the plane of scattering are identical, which rules out the existence of birefringent effects.

\subsection{Significance}

There are some quick and general conclusions that can be drawn from this calculation. One is that massless particles no longer move along null geodesics. We have obtained the classical behavior as a first approximation, and indeed we even have obtained the correct result to second order. However there are new effects the interaction which have a different power dependence which will then modify the trajectory. Moreover, some of these interactions are different depending on the type of massless particle, a scalar vs a photon. This is a violation of some classical forms of the equivalence principle. The equivalence principle itself means different things in different settings\cite{ssy}, yet classically would state that massless particles move on null geodesics which are the same for all massless particles. Both the geodesic motion and the universality turn out to be violated by quantum corrections. Both of these effects can be ascribed to the long distance propagation of particles in loops. These loop diagrams then sample spacetime points different from the classical geodesic. Particles have intrinsic power-law nonlocality and this leads to non-classical motion.

The most familiar application of the scattering of massless and massive systems is probably the bending of starlight by the sun.  A fully quantum treatment of this light bending which is capable of including the one-loop amplitude effects is not available.
However, in order to try to understand the impact of these corrections, one can proceed by defining, in the small momentum transfer limit $t\simeq -{\boldsymbol q}^{\,2}$, a semi-classical potential for a massless scalar and photon interacting with a massive scalar object by use of the Born approximation result
\vskip-0.6cm
\begin{eqnarray}
V_\eta(r)\!\!&=&\!\!{\hbar\over 4M\omega}\!\! \int\!\!
\cM^{[\eta(p_1)\eta(p_2)]}_{[\phi(p_3)\phi(p_4)]}(\boldsymbol q)\, e^{i \boldsymbol
  q\cdot r}\! {d^3q\over (2\pi)^3}\cr
&\hspace{-0.8cm}\simeq& -{2G M \omega\over r}+{15\over 4}{(GM)^2 \omega\over r^2}\cr
&+& {8 bu^\eta-15\over 4\pi}\,{G^2 M \omega\hbar\over r^3}
+ {12 G^2 M \omega\hbar\over \pi}\,{\log{r\over r_0}\over r^3}\,.
\end{eqnarray}
\vskip-0.1cm\noindent
where $r_0$ is an infrared scale.

Using na\"ively the semi-classical formula for angular deflection given in~\cite[chap.~21]{bhm}--\cite{dya} and the above potential we find the bending angle of a photon and for a massless scalar
\vspace{-0.15cm}
\begin{eqnarray}\label{e:theta}
&&\theta_\eta \simeq- {b\over \omega}\int_{-\infty}^{+\infty} {V_\eta'(b \sqrt{1+u^2})\over
  \sqrt{1+u^2}} du\\
\nonumber &\simeq& {4G M\over b}+{15\over 4} {G^2 M^2 \pi \over
  b^2}
+{8 bu^\eta+9+48 \log {b\over 2r_o}\over \pi}\,{G^2 \hbar M\over  b^3}\,.
\end{eqnarray}
\vskip-0.15cm\noindent
The first two terms give the well known classical values, including the first post-Newtonian correction, expressed in term of the gauge-invariant impact parameter $b$ (see for instance~\cite{wgl}).  The last term is a quantum gravity effect of order $G^2\hbar M/b^3=\ell_P^2r_S/(2b^3)$ and involves the product of the Planck length and the Schwarzschild radius of the massive object divided by the cube of the impact parameter and depends on the spin of massless particle scattering on the massive target.  Of course, this dependence does {\it not} necessarily violate the equivalence principle in the most fundamental sense, in that the logarithmic quantum corrections correspond to non-local effects in coordinate space.  Because of quantum loop effects, the long-distance propagation of massless photons and gravitons is not localized, and consequently can be interpreted as a tidal correction in that the massless particle is no longer be describable as a point source.  There is then no requirement from the equivalence principle that such non-local effects be independent of the spin of the massless particle.

Numerically, we can compare the bending angle of a photon with that of a massless scalar by the sun.  The only difference
given the above treatment is given by the massless bubble effect
\begin{equation}
  \theta_\gamma-\theta_\varphi =  {8 (bu^\gamma-bu^\varphi)\over \pi} \,
  {G^2\hbar M\over b^3}  \,.
\end{equation}
and is far too small to be seen experimentally~\cite{wva}.  However, it is interesting that quantum effects {\it do} predict such a difference, without any free parameter, modifying one of the key features of classical general relativity.  Moreover, this phenomenon represents another demonstration that effective field techniques can make well-defined predictions within quantum gravity.

\section{Massless particle scattering}

There are also calculations of the gravitational scattering involving all massless particles. These include a remarkable calculation of graviton-graviton scattering by Dunbar and Norridge\cite{dbr}, a calculation of the scattering of a real scalar by the same authors \cite{drb} and the scattering of two non-identical scalars\cite{abr} obtained using the methods of \cite{drb}. The character of the results are similar to the reactions discussed above, so we can refer the reader to the original papers for the results. However, we do want to comment on a couple of features of the massless amplitudes.

One interesting feature is that there are no ``classical'' corrections in the massless amplitudes. We have seen this partially in the previous section, where the square-root non-analyticity is associated with the massive leg only not with the massless field. Amplitudes of totally massless particle only involve logarithms. Therefore the loop expansion is here strictly the $\hbar$ expansion, and one does not build up classical gravitational solutions in the intermediate states.

A second interesting feature is the ubiquity of infrared singularities and non-local effects. All of these scattering amplitudes can be decomposed into coefficients times the bubble, triangle and box diagrams. For massless particles these are particularly simple:
\begin{eqnarray}
I_2(s) &=& r_\Gamma\left[ \frac{1}{\epsilon} + 2 -\ln (-s)~\right]\nonumber\\
I_3(s) &=&    -  r_\Gamma   \frac{1}{s}\left[\frac{1}{\epsilon^2} - \frac{\ln (-s)}{\epsilon} +\frac12 \ln^2 (-s) \right]   \nonumber \\
I_4(s,t) &=&  r_\Gamma   \frac{1}{st}\left[\frac{4}{\epsilon^2} - 2\frac{\ln(-s) ~+\ln)-t)}{\epsilon} +2\ln(-s)~\ln(-t) -\pi^2\right]
\end{eqnarray}
with
\begin{equation}
r_\Gamma = \frac{\Gamma^2(1-\epsilon) \Gamma(1+\epsilon)}{\Gamma(2-\epsilon)}    \ \ .
\end{equation}
Of the many divergences that are displayed in these equations, only the $1/\epsilon$ in the bubble diagram $I_2(s)$ is of ultraviolet origin. Therefore only it and the related $+2$ in the same amplitude represent local physics. All the other $1/\epsilon$ term in the other amplitudes represent infrared divergences, and the logarithms represent non-local effects. This tells us that almost all loop processes in gravity are either infrared divergent or non-local. While these
features are well understood in scattering amplitudes, they are far from well understood in other gravitational settings such as cosmology and classical solutions. The IR portions of loops point to new phenomena in gravitational physics.

\section{Conclusion}

Although a renormalizable theory which merges general relativity and quantum mechanics has yet to be identified by experiment, we have shown above that when treated
as a nonrenormalizable effective field theory, quantum gravity is a very successful theory.  An EFT represents an expansion in powers of derivatives (energy-momentum) divided by a scale parameter, needed to make the expansion parameter dimensionless.  In the case of chiral perturbation theory, which is a very successful picture of low energy QCD, the chiral scale parameter $\Lambda_\chi\sim 4\pi F_\pi\sim 1$ GeV so that $\chi$pt is valid at energies $\sim\leq$ 500 MeV.  On the other hand, the scale parameter in the case of quantum gravity is $\Lambda_{grav}\sim M_{pl}\sim 10^{19}$ GeV so that effective gravity is valid at any energy which is reachable in current accelerators.  This means that higher order gravitational counterterms have essentially no influence {\it on} current experiments.  On the other hand, the existence of gravitational loop effects means that various non-analytic terms such as $\sqrt{-q^2\over m^2}$ or $\log{-q^2\over m^2}$ are present in transition amplitudes, indicating, after Fourier transform, the presence of long distance---$1/r^n$ with $n=2,3,4...$---effects in quantities such as the energy-momentum tensor, the metric tensor, the interaction potential, etc.  Despite the fact that these are loop effects, they are expected from classical physics arguments and in the case of $T_{\mu\nu}$ and $h_{\mu\nu}$ these forms agree with well known classical solutions.  In addition loop effects produce quantum mechanical corrections to the classical results of order $\hbar/(mr)^{n+1}$ with $n=2,3,4...$. Indeed, it is the nonlocal quantum effects which are the purest manifestation of the EFT.

We have evaluated a series of scattering amplitudes and focused on the parts of the calculation that the effective theory is capable of calculating.
We found in each case that the quantum corrected amplitude is well defined when gravity is treated as an effective field theory.
Some of the results that we displayed were:
\begin{itemize}
\item[i)] Gravitational corrections to the energy-momentum tensor of a massive system:  one loop gravitational corrections were calculated for matrix elements of the energy-momentum tensor $<p_2|T_{\mu\nu}(x)|p_1>$.  The results were found to agree with the classical forms of the gravitational energy-momentum tensor for both spinless and spin 1/2 systems.  Converting to the metric tensor by use of the (linear) Einstein equation, the spinless case was found to agree with the Schwarzschild solution, while in the case of spin 1/2 the corrections were shown to match the Kerr solution.  Quantum corrections to these classical results were determined in both cases.\\

\item[ii)] Gravitational scattering of two massive particles:  one-loop gravitational corrections to the potential which characterizes the interaction of particles with mass $m_1$ and $m_2$ were evaluated.  At lowest order the potential is simply the classic Newtonian result.  However, at higher order modifications are found and once again there exist both long distance classical {\it and} quantum mechanical corrections.  Here we found a universal soft theorem, such that the form of the quantum correction is universal, coming as a reflection of the soft theorems of tree amplitudes.\\

\item[iii)] Gravitational scattering of a massive and massless particle:  one loop gravitational corrections were calculated to the interaction of a massive test particle with a massless scalar and photon.  In the case of a Feynman diagram calculation involving a massless system the higher order energy-momentum tensor is not well defined and the calculation of the scattering amplitude requires the use of both photon and graviton mass regulators.  Individual diagrams have numerous divergences but when the total is calculated, there exists a subtle cancelation of the various divergences so that the final result is finite.  However, modern on-shell helicity amplitude methods were shown to provide a significantly simplified route to this result. Here we found that
    massless particles no longer follow null geodesics, and that different types of massless particles have different trajectories. Both of these results are deviations from classical behavior.
\item[iv)] Although we have not highlighted this in the discussion above, one can also see from these results that there is not a form of a ``running'' coupling $G(E)$ in the effective theory\cite{abr}. There is no universality of the quantum corrections which could have been absorbed into a running coupling. This result is totally expected in the effective field theory. However it is worth stating, as there are many attempts in the literature to define such a running $G$.

\end{itemize}

Of course, general relativity is not primarily concerned with scattering amplitudes. We have started with these because scattering is what perturbative quantum field theory does best. However, the next challenge becomes to extend these effective field theory techniques to other solutions of general relativity. The metric
calculations described above are a start down this path. There have also been applications to cosmology\cite{bsm}. But much remains to be done to understand the low energy quantum predictions that can be calculated in the effective field theory.

\begin{center}
{\bf Acknowledgment}
\end{center}

The work of JFD is partially supported by the National Science Foundation under award PHY-1205986.

\begin{center}
{\bf Appendix A}
\end{center}

In this section we derive the various couplings to be used in our calculation.  We begin with the scalar field, whose
matter action is
\begin{equation}
S_m=\int d^4x\sqrt{-g}{1\over 2}\left(g^{\mu\nu}D_\mu\phi D_\nu\phi-m^2\phi^2\right)
\end{equation}
where $D_\mu$ is the covariant derivative with respect to the background field.  (In our case quantizing about flat space we have $D_\mu=\partial_\mu$.)

The gravitational coupling of a spin-0 particle is found by expanding
the minimally coupled scalar field matter Lagrangian
\begin{equation}
\sqrt{-g}{\cal L}_m=\sqrt{-g}\left({1\over 2} g^{\mu \nu} \partial_\mu \phi
\partial_\nu \phi  -{1\over 2}m^2\phi^2 \right)
\end{equation}
in terms of the gravitational field $h_{\mu\nu}$ which is a small
fluctuation of the metric about flat Minkowski space defined as
\begin{equation}
g_{\mu\nu}=\eta_{\mu\nu} + \kappa h_{\mu\nu}^{(1)}
\end{equation}
with $\kappa = \sqrt{32 \pi G} \propto 1/M_P$. The inclusion of this
factor $\kappa$ in the definition of the graviton field $h_{\mu
\nu}$ gives this field  a mass-dimension of unity and thus yields a
kinetic term of standard normalization without a dimensionful
parameter. For matter interactions, this choice is convenient since
the order of $\kappa$ keeps track of the number of gravitons
involved in an interaction. Once the action is written in terms of
the expansion of the graviton field, all indices are understood to
be lowered or raised using the Minkowski metric $\eta_{\mu \nu}$. We
also require the expansion of the inverse metric and square root of
the determinant of the metric tensor---
\begin{eqnarray}
g^{\mu\nu}&=&\eta^{\mu\nu} - \kappa h^{(1)\mu\nu} + \kappa^2 h^{(1)\mu\alpha} {h^{(1)\nu}}_\alpha
 + \mathcal O(\kappa^3)\nonumber \\
\sqrt{-g}&=& 1 + \frac{\kappa}{2} h^{(1)} + \frac{\kappa^2}{8} \left(h^{(1)2} - 2 h_{\mu \nu}^{(1)}
h^{(1)\mu \nu} \right) + \mathcal O(\kappa^3).
\end{eqnarray}
Then, expanding in powers of $\kappa$, we find
\begin{eqnarray}
\sqrt{-g}{\cal L}_m^{(0)}&=&{1\over
2} \partial_\mu\phi\partial^\mu\phi - \frac{1}{2} m^2\phi^2 \nonumber\\
\sqrt{-g}{\cal L}_m^{(1)}&=&{\kappa\over 2} \, h^{(1)\mu\nu}\left[\eta_{\mu\nu}
\left(\frac{1}{2} \partial_\alpha\phi\partial^\alpha\phi - \frac{1}{2} m^2\phi^2\right)-
\partial_\mu\phi\partial_\nu\phi\right]\nonumber\\
\sqrt{-g}{\cal L}_m^{(2)}&=&{\kappa^2\over 2} \hspace*{0.5pt} \Bigg[\frac{1}{4} \Big(h^{(1)2}
- 2 h_{\mu \nu}^{(1)} h^{(1)\mu\nu}\Big) \bigg(\frac{1}{2} \partial_\alpha \phi \partial^\alpha \phi
- \frac{1}{2} m^2 \phi^2\bigg)\nonumber\\
&& \hspace*{13pt} + \bigg(h^{(1)\mu \alpha} {h^{(1)\nu}}_\alpha - \frac{1}{2} h^{(1)} h^{(1)\mu\nu}\bigg)
\partial_\mu \phi \partial_\nu \phi \Bigg]
\end{eqnarray}
where $h^{(1)}\equiv \eta^{\alpha\beta}h_{\alpha\beta}^{(1)}$ represents the
trace so the one- and two-graviton vertices are identified as\\
\begin{figure}[ht]
  \centering
  \includegraphics{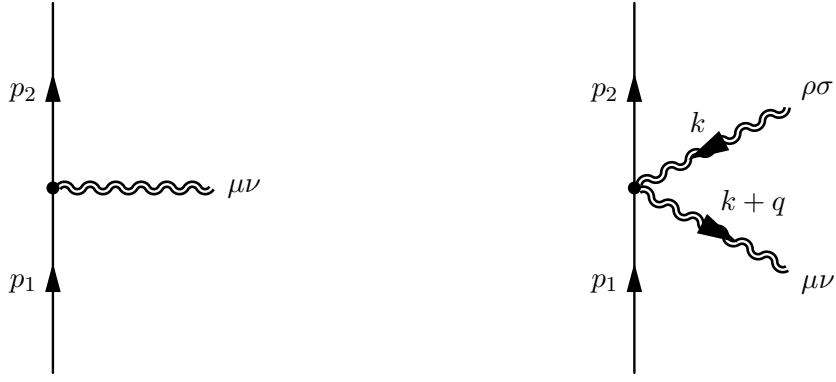}
  \caption{The one- and two-graviton couplings}
\end{figure}
\vspace*{-20pt}
\begin{eqnarray}
{}^{0} \tau_{\mu\nu}^{(1)}(p_2, p_1, m) \hspace*{-7.5pt}&=\hspace*{-7.5pt}& {-i \kappa\over
2}\left[p_{1\mu}p_{2\nu}+p_{1\nu}p_{2\mu}-\eta_{\mu\nu}(p_1\cdot
p_2-m^2)\right]\nonumber\\
{}^{0} \tau^{(2)}_{\mu\nu,\rho\sigma}(p_2, p_1, m)\hspace*{-7.5pt}&=\hspace*{-7.5pt}&
\frac{i\kappa^2}{2}\bigg[2 I_{\mu\nu,\kappa\zeta}{I^\zeta}_{\lambda,\rho\sigma}(p_1^\kappa
p_2^\lambda+p_1^\lambda p_2^\kappa) \nonumber\\
&& {}\hspace*{14pt} - \hspace*{-1pt} (\hspace*{-0.5pt} \eta_{\mu\nu}I_{\kappa\lambda,\rho\sigma} \hspace*{-1.5pt} + \hspace*{-1.5pt} \eta_{\rho\sigma}I_{\kappa\lambda,\mu\nu}\hspace*{-0.5pt} )p_1^\kappa p_2^\lambda \hspace*{-1pt}
- \hspace*{-1pt} P_{\mu\nu, \rho\sigma} (p_1 \hspace*{-1.5pt} \cdot \hspace*{-1.3pt} p_2 \hspace*{-1.8pt} - \hspace*{-1.5pt} m^2) \hspace*{-1pt} \bigg]\nonumber\\
\quad  \label{eq:tp}
\end{eqnarray}
where we have defined
\begin{eqnarray}
I_{\alpha\beta,\gamma\delta}&\equiv&{1\over
2}(\eta_{\alpha\gamma}\eta_{\beta\delta}+\eta_{\alpha\delta}\eta_{\beta\gamma}) \nonumber \\
P_{\alpha\beta,\gamma\delta}&\equiv&{1\over
2}(\eta_{\alpha\gamma}\eta_{\beta\delta}+\eta_{\alpha\delta}\eta_{\beta\gamma} - \eta_{\alpha\beta}\eta_{\gamma\delta}).
\end{eqnarray}

\begin{figure}
\begin{center}
  \epsfig{file=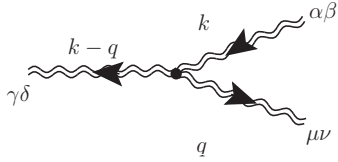,width=5cm} \caption{ The three
  graviton vertex }
\end{center}
\end{figure}

\medskip
\medskip

We also require the energy-momentum tensor for gravitons which leads to the triple-graviton vertex\cite{jfd}
\begin{eqnarray}
\tau^{\mu\nu}_{\alpha\beta,\gamma\delta}(k,q)&=&{i\kappa\over 2}\left\{
P_{\alpha\beta,\gamma\delta}
\left[k^\mu k^\nu+(k-q)^\mu (k-q)^\nu+q^\mu q^\nu-{3\over
2}\eta^{\mu\nu}q^2\right]\right.\nonumber\\
&+&\left.2q_\lambda
q_\sigma\left[I^{\lambda\sigma,}{}_{\alpha\beta}I^{\mu\nu,}
{}_{\gamma\delta}+I^{\lambda\sigma,}{}_{\gamma\delta}I^{\mu\nu,}
{}_{\alpha\beta}-I^{\lambda\mu,}{}_{\alpha\beta}I^{\sigma\nu,}
{}_{\gamma\delta}-I^{\sigma\nu,}{}_{\alpha\beta}I^{\lambda\mu,}
{}_{\gamma\delta}\right]\right.\nonumber\\
&+&\left.[q_\lambda
q^\mu(\eta_{\alpha\beta}I^{\lambda\nu,}{}_{\gamma\delta}
+\eta_{\gamma\delta}I^{\lambda\nu,}{}_{\alpha\beta})+ q_\lambda
q^\nu(\eta_{\alpha\beta}I^{\lambda\mu,}{}_{\gamma\delta}
+\eta_{\gamma\delta}I^{\lambda\mu,}{}_{\alpha\beta})\right.\nonumber\\
&-&\left.q^2(\eta_{\alpha\beta}I^{\mu\nu,}{}_{\gamma\delta}+\eta_{\gamma\delta}
I^{\mu\nu,}{}_{\alpha\beta})-\eta^{\mu\nu}q^\lambda
q^\sigma(\eta_{\alpha\beta}
I_{\gamma\delta,\lambda\sigma}+\eta_{\gamma\delta}
I_{\alpha\beta,\lambda\sigma})]\right.\nonumber\\
&+&\left.[2q^\lambda(I^{\sigma\nu,}{}_{\alpha\beta}
I_{\gamma\delta,\lambda\sigma}(k-q)^\mu
+I^{\sigma\mu,}{}_{\alpha\beta}I_{\gamma\delta,\lambda\sigma}(k-q)^\nu\right.\nonumber\\
&-&\left.I^{\sigma\nu,}{}_{\gamma\delta}I_{\alpha\beta,\lambda\sigma}k^\mu-
I^{\sigma\mu,}{}_{\gamma\delta}I_{\alpha\beta,\lambda\sigma}k^\nu)\right.\nonumber\\
&+&\left.q^2(I^{\sigma\mu,}{}_{\alpha\beta}I_{\gamma\delta,\sigma}{}^\nu+
I_{\alpha\beta,\sigma}{}^\nu
I^{\sigma\mu,}{}_{\gamma\delta})+\eta^{\mu\nu}q^\lambda q_\sigma
(I_{\alpha\beta,\lambda\rho}I^{\rho\sigma,}{}_{\gamma\delta}+
I_{\gamma\delta,\lambda\rho}I^{\rho\sigma,}{}_{\alpha\beta})]\right.\nonumber\\
&+&\left.[(k^2+(k-q)^2)\left(I^{\sigma\mu,}{}_{\alpha\beta}I_{\gamma\delta,\sigma}{}^\nu
+I^{\sigma\nu,}{}_{\alpha\beta}I_{\gamma\delta,\sigma}{}^\mu-{1\over
2}\eta^{\mu\nu}P_{\alpha\beta,\gamma\delta}\right)\right.\nonumber\\
&-&\left.(k^2\eta_{\gamma\delta}I^{\mu\nu,}{}_{\alpha\beta}+(k-q)^2\eta_{\alpha\beta}
I^{\mu\nu,}{}_{\gamma\delta})]\right\}
\end{eqnarray}
\vskip-0.1cm\noindent

For the case of spin-1/2 we require some additional formalism in
order to extract the gravitational couplings, which is necessary
because the Dirac algebra $\big\{\gamma^a , \gamma^b \big\} = 2 \eta^{a b}$
is defined with respect to the Minkowski flat space metric.
In this case the Dirac matter Lagrangian coupled to gravity reads
\begin{equation}
\sqrt{-g}{\cal L}_m = \sqrt {-g} \, \bar \psi \left[\frac{i}{2} \hspace{1pt} {e^{\mu}}_a \{ \gamma^a, D_\mu \} - m\right]\psi
\end{equation}
and involves the vierbein ${e^{\mu}}_a$ which links global
coordinates with those in a locally flat space.  The vierbein is
in some sense the ``square root'' of the metric tensor
$g_{\mu\nu}$ and satisfies the relations
\begin{eqnarray}
{e_{\mu}}^a \hspace*{1pt} {e_{\nu}}^b \, \eta_{ab} &=& g_{\mu \nu} \quad \quad \quad \quad {e^{\mu}}_a \hspace*{1pt} {e^{\nu}}_b \, \eta^{ab} \ \, = \, \ g^{\mu \nu} \nonumber\\
{e_{\mu}}^a \hspace*{1pt} {e_{\nu}}^b \, g^{\mu \nu} &=& \eta_{ab} \hspace*{0.5pt} \quad \quad \quad \quad {e^{\mu}}_a \hspace*{1pt} {e^{\nu}}_b \, g_{\mu \nu} \ \, = \ \, \eta^{ab} .
\end{eqnarray}
The covariant derivative is
\begin{equation}
 D_\mu = \frac{1}{2} \hspace{1pt} \partial^{LR}_\mu + \frac{i}{4} \, {{\omega_{\mu}}^a}_b \, \eta_{a c} \ \sigma^{c b}
\end{equation}
with $\sigma^{c b} = \frac{i}{2} \Big[\gamma^c, \gamma^d\Big]$ and the partial derivative $\partial^{LR}_\mu$ acts only on spinors and in such a way that
\begin{equation}
 \bar \psi \partial^{LR}_\mu \psi = \bar \psi \, \partial_\mu \psi - \left(\partial_\mu \bar \psi\right) \psi.
\end{equation}
Putting everything together, we find then
\begin{equation} \label{lagrangian12}
 \sqrt {-g} \, \mathcal L_m = \sqrt {-g} \, \bar \psi \left[\frac{i}{2} \hspace{1pt} \gamma^a {e^{\mu}}_a \partial^{LR}_\mu - \frac{1} {8} \hspace{1pt} {e^{\mu}}_{a'} \ {{\omega_{\mu}}^a}_b \, \eta_{a c} \, \{\gamma^{a'} , \sigma^{c b}\} - m\right]\psi.
\end{equation}
The spin connection ${{\omega_{\mu}}^a}_b \, \eta_{a c}$ can be
derived in terms of vierbeins by requiring $D_\mu {e_\nu}^a = 0$ and
by antisymmetrization in $\mu \leftrightarrow \nu$ in order to get
rid of Christoffel symbols\footnote{For our purposes we shall use
only the symmetric component of the vierbein matrices, since these
are physical and can be connected to the metric tensor, while their
antisymmetric components are associated with freedom of homogeneous
transformations of the local Lorentz frames and do not contribute to
nonanalyticity\cite{dsr}}. The result is:
\begin{equation}
 {{\omega_{\mu}}^a}_b \, \eta_{a c} = \left(\frac{\eta_{ab}} {2} \, {e^\nu}_c \left(\partial_\mu {e_\nu}^a - \partial_\nu {e_\mu}^a \right) + \frac{\eta_{af}}{2} \, {e^\nu}_c \hspace{1pt} {e^\rho}_b \hspace{1pt} {e_\mu}^f \hspace{1pt} \partial_\rho {e_\nu}^a \right) - \Big(b \leftrightarrow c\Big)
\end{equation}
In order to derive the Feynman rules we expand the ingredients in
Eq. (\ref{lagrangian12}) that contain graviton couplings, that is we
need ${e^{\mu}}_a$ and ${{\omega_{\mu}}^a}_b \, \eta_{a c}$ expanded
up to $\mathcal O(\kappa^2)$
\begin{eqnarray}
 {e_\mu}^a \hspace*{-3pt}& = \hspace*{-3pt}&\delta_\mu^a + \frac{\kappa}{2} \hspace{1pt} {h^{(1)a}}_\mu
  - \frac{\kappa^2}{8} \hspace{1pt} h_{\mu \rho}^{(1)} h^{(1)a\rho}+\ldots \nonumber \\
 {e^\mu}_a \hspace*{-3pt}& = \hspace*{-3pt}& \delta_a^\mu - \frac{\kappa}{2} \hspace{1pt} {h^{(1)\mu}}_a
  + \frac{3 \kappa^2}{8} \hspace{1pt} h_{a \rho}^{(1)} h^{(1)\mu\rho}+\ldots \nonumber \\
 {{\omega_{\mu}}^a}_b \, \eta_{a c} \hspace*{-3pt}& = \hspace*{-3pt}&\frac{\kappa}{2}
 \hspace{1pt} \partial_b h_{\mu c}^{(1)} + \frac{\kappa^2}{8} \hspace{1pt} {h^{(1)\rho}}_b \partial_\mu h_{c \rho}^{(1)}
 - \frac{\kappa^2}{4} \hspace{1pt} {h^{(1)\rho}}_b \partial_\rho h_{\mu c}^{(1)} + \frac{\kappa^2}{4} \hspace{1pt} {h^{(1)\rho}}_b \partial_c h_{\mu \rho}^{(1)} - \Big(b \leftrightarrow c\Big) \nonumber \\
\end{eqnarray}
After these expansions are employed, we no longer need to
distinguish between Latin Lorentz ind
ices and Greek covariant
indices and can use the Minkowski metric to lower and raise all indices.

The matter Lagrangian then has the expansion---(note here that our
conventions are $\gamma_5=-i\gamma^0\gamma^1\gamma^2\gamma^3$ and
$\epsilon^{0123} = + 1$)
\begin{eqnarray}
\sqrt{-g}{\cal L}_m^{(0)}&=&\bar{\psi} \hspace*{-1pt} \left({i\over 2} \hspace*{-2pt} \not\!{\partial}^{LR}-m \right) \hspace*{-1.3pt} \psi\nonumber\\
\sqrt{-g}{\cal L}_m^{(1)}&=&{\kappa\over 2} \hspace*{1pt} h^{(1)} \, \bar{\psi} \hspace*{-1pt}\left({i\over 2} \hspace*{-2pt} \not\!{\partial}^{LR}-m\right) \hspace*{-1.3pt}\psi
  -{\kappa\over 2} \hspace*{1pt} h^{(1)\mu \nu} \, \bar{\psi} \, \frac{i}{2} \partial_\mu^{LR} \gamma_\nu \hspace*{1pt} \psi\nonumber\\
\sqrt{-g}{\cal L}_m^{(2)}&=& {\kappa^2 \over 8} \left(h^{(1)2} - 2 h_{\alpha \beta}^{(1)} h^{(1)\alpha \beta} \right) \bar{\psi} \hspace*{-1pt} \left({i\over 2} \hspace*{-2pt} \not\!{\partial}^{LR}-m\right) \hspace*{-1.3pt} \psi \nonumber \\
&+& \frac{\kappa^2}{8} \left(3 h^{(1)\mu \rho} {h^{(1)\nu}}_\rho - 2 h^{(1)} h^{(1)\mu\nu} \right) \bar{\psi} \, \frac{i}{2} \partial_\mu^{LR} \gamma_\nu \hspace*{1pt} \psi\nonumber\\
&+& \frac{i \kappa^2}{16}  \epsilon^{\alpha \beta \gamma \delta} \,{ h^{(1)\rho}}_\alpha (i\partial_\beta h_{\rho \gamma}^{(1)}) \, \bar \psi \gamma_\delta \gamma_5 \psi
\end{eqnarray}
and the corresponding one- and two-graviton vertices are found to be

\begin{eqnarray}
{}^{\frac{1}{2}}\tau_{\mu\nu}^{(1)}(p_2, p_1, m)\hspace*{-8pt} & = \hspace*{-7.5pt}&{-i\kappa\over 2}\hspace*{-0.5pt}\Bigg[\hspace*{-0.5pt}{1\over 4}\hspace*{-0.7pt}\Big(\hspace*{-1pt}\gamma_\mu(p_1 \hspace*{-1.8pt} + \hspace*{-1.5pt} p_2)_\nu \hspace*{-1pt}+\hspace*{-1pt}\gamma_\nu(p_1 \hspace*{-1.8pt} + \hspace*{-1.5pt} p_2)_\mu\hspace*{-0.5pt}\Big)
\hspace*{-2.3pt} - \hspace*{-1.5pt} \eta_{\mu\nu} \hspace*{-0.7pt} \Bigg(\hspace*{-1pt}{1\over 2}(\! \not\!{p}_1 + \hspace*{-1pt} \!\not\!{p}_2)\hspace*{-2pt}- \hspace*{-1pt} m \hspace*{-1.5pt} \Bigg) \hspace*{-2pt} \Bigg]\nonumber\\
{}^{\frac{1}{2}}\tau_{\mu\nu,\rho\sigma}^{(2)}(p_2, p_1, m) \hspace*{-8pt} & = \hspace*{-7.5pt}& i\kappa^2\Bigg[
\hspace*{-1pt}-{1\over 2}\bigg({1\over 2}(\not \!{p}_1 \hspace*{1pt} + \! \not\!{p}_2)-m\bigg) \hspace*{1pt} P_{\mu\nu,\rho\sigma}\nonumber\\
&& \hspace*{20.3pt}{} - {1\over
16}\bigg[\eta_{\mu\nu} \Big(\gamma_\rho(p_1+p_2)_\sigma+\gamma_\sigma(p_1+p_2)_\rho\Big) \nonumber\\
&& \hspace*{43.5pt} {} + \hspace*{-1pt} \eta_{\rho\sigma}\Big(\gamma_\mu
(p_1+p_2)_\nu+\gamma_\nu(p_1+p_2)_\mu\Big)\bigg] \nonumber\\
&& \hspace*{20.3pt}{} + {3\over
16}(p_1+p_2)^{\epsilon}\rho^{\xi}(I_{\xi\phi,\mu\nu}{I^{\phi}}_{\epsilon,\rho\sigma}
+I_{\xi\phi,\rho\sigma}{I^{\phi}}_{\epsilon,\mu\nu}) \nonumber\\
&& \hspace*{20.3pt}{} + {i\over 16}\epsilon^{\epsilon\phi\eta\lambda}\gamma_\lambda \gamma_5
\Big(I_{\rho\sigma,\phi\xi} {I_{\mu\nu,\eta}}^\xi \hspace*{1pt} {k}_\epsilon
- I_{\mu\nu,\phi\xi} {I_{\rho\sigma,\eta}}^\xi \hspace*{1pt} (k+q)_\epsilon \Big)\Bigg]. \nonumber \\ \quad
\end{eqnarray}
\begin{center}
{\bf Appendix B}
\end{center}
Above we have argued that the scattering amplitude which is
ultimately related to observables in quantum field theory is a
physical quantity while the potential we have given is not
and depends on the gauge, the choice of
coordinates, and on the way the iteration is performed, {\it
i.e.}, on the way we do the matching. While the classical
component of our potential is in fact plagued by these ambiguities,
the quantum part is unique since it is unaffected by how we perform
the matching and a quantum field theory calculation in any
gauge would result in the same result\cite{bdh}.

In this appendix we demonstrate how we can recover the classical equations of motion
from our scattering amplitudes by setting up the Einstein-Infeld-Hoffmann (EIH) Lagrangian\cite{eih}.
The EIH Lagrangian is itself dependent on the choice of coordinates,
but can be expressed in the center of mass frame
($\boldsymbol P \equiv \boldsymbol p_1 = - \boldsymbol p_3$, $\boldsymbol r \equiv \boldsymbol r_1 - \boldsymbol r_3$) in a general way as\cite{bhi}
\begin{equation}
 L_{EIH} = T - V
\end{equation}
where the kinetic energy to NLO in the nonrelativistic expansion reads
\begin{equation}
 T = \frac{\boldsymbol P^{\hspace*{1.4pt} 2}}{2 m_1} + \frac{\boldsymbol P^{\hspace*{1.4pt} 2}}{2 m_2} - \frac{\boldsymbol P^{\hspace*{1.4pt} 4}}{8 m_1^3} - \frac{\boldsymbol P^{\hspace*{1.4pt} 4}}{8 m_2^3}
\end{equation}
and the potential is
\begin{equation}
 V = V^{(1)} + V^{(2)}
\end{equation}
with
\begin{eqnarray}
 V^{(1)} & = & - \frac{G m_1 m_2}{r} \left\{1 + \left[\frac{1}{2} + \left(\frac{3}{2} - \alpha\right) \frac{(m_1 + m_2)^2}{m_1 m_2}\right] \frac{\boldsymbol P^{\hspace*{1.4pt} 2}}{m_1 m_2} \right. \nonumber \\
  && \hspace*{69.5pt} \left.+ \left[\frac{1}{2} + \alpha \, \frac{(m_1 + m_2)^2}{m_1 m_2} \right] \frac{\left(\boldsymbol P \cdot \hat r\right)^2}{m_1 m_2}\right\} \label{eq_eihV1} \\
 V^{(2)} & = & \left(1 - 2 \alpha\right) \frac{G^2 m_1 m_2 (m_1 + m_2)}{2 r^2}. \label{eq_eihV2}
\end{eqnarray}
The parameter $\alpha$ parameterizes the choice of coordinates used, with $\alpha = 0$ being the gauge of the original EIH result.
The coordinate change
\begin{equation}
 \boldsymbol r \rightarrow \boldsymbol r \hspace*{1pt} \left(1 - \alpha \hspace*{1pt} \frac{G (m_1 + m_2)}{r}\right)
\end{equation}
which implies
\begin{equation}
 \boldsymbol P \rightarrow \boldsymbol P + \alpha \hspace*{1pt} \frac{G (m_1 + m_2)}{r} \left[\boldsymbol P - \left(\boldsymbol P \cdot \hat r \right) \hat r \right]
\end{equation}
then brings the original EIH Lagrangian into the form above, which is the
most general result.

Since we perform our matching on-shell, {\it i.e.}, we use the
on-shell one-graviton exchange amplitude to define the leading order
$\mathcal O(G)$ potential, terms proportional to $\boldsymbol P \cdot \hat
r$ never arise, meaning that our result corresponds to a gauge such
that the coefficient of the structure
$$\frac{\left(\boldsymbol P \cdot \hat r\right)^2}{m_1 m_2}$$ in Eq. (\ref{eq_eihV1}) vanishes.
That is the case if and only if the gauge parameter is
\begin{equation}
 \alpha = - \frac{m_1 m_2}{2(m_1 + m_2)^2}
\end{equation}
whereby the EIH potential becomes
\begin{eqnarray}
 V^{(1)} & = & - \frac{G m_1 m_2}{r} \left\{1 + \left[1 + \frac{3}{2} \hspace*{0.8pt} \frac{(m_1 + m_2)^2}{m_1 m_2}\right] \frac{\boldsymbol P^{\hspace*{1.4pt} 2}}{m_1 m_2} \right\} \label{eq_eihHRV1} \\
 V^{(2)} & = & \left(1 + \frac{m_1 m_2}{(m_1 + m_2)^2}\right) \frac{G^2 m_1 m_2 (m_1 + m_2)}{2 r^2}. \label{eq_eihHRV2}
\end{eqnarray}
Comparing the EIH potential $V^{(1)}$ in this gauge of Eq.
(\ref{eq_eihHRV1}) with the long distance component of the leading
order spin-independent potential in Eq. (\ref{eq:po}) we find full
agreement for the relativistic corrections to the $\mathcal O(G)$
potential. However, comparing the EIH potential $V^{(2)}$ in this
gauge---Eq. (\ref{eq_eihHRV2})---with the classical component of our
spin-independent potential in Eq. (\ref{eq:so}) we see that the two
do {\it not} agree!  The reason for this discrepancy is that we elected to
use nonrelativistic forms when we performed the second Born
iteration of the leading order potential in Eq. (\ref{eq:ib}).  This procedure, however, is not
self-consistent when we are interested in equations of motion at
NLO, and we must account for the leading relativistic
corrections in performing the iteration.  In particular, we must use
expressions for the potential and the propagator in Eq. (\ref{eq:ib}) which include these leading relativistic
terms\footnote{The subscript $NLO$ in this sections refers to
the iteration being performed at NLO in the relativistic expansion.}
\begin{equation}
\left<\boldsymbol p_f \left| {}^0 \hat V^{(1)}_{NLO} \right| \boldsymbol p_i
\right> \simeq -{4\pi Gm_1m_2\over
\boldsymbol{q}^{\hspace*{1.4pt}2}}\left[1+{\boldsymbol p_i^{\hspace*{1.4pt}2} +
\boldsymbol p_f^{\hspace*{1.4pt}2} \over 2m_1m_2}\left(1 + \frac{3(m_1 + m_2)^2}{2 m_1 m_2}\right)\right]
\end{equation}
\begin{equation}
 G_{NLO}^{(0)}(\ell)={i\over {p_0^2\over 2m_r}-{\ell^2\over 2m_r}+i\epsilon} \times \left[1 + \left({p_0^2\over 4m_r^2}+{\ell^2\over 4m_r^2}\right) \left(1-3{m_r^2\over m_1m_2}\right)\right]
\end{equation}
which yields a second Born iteration amplitude
\begin{eqnarray}
{}^0{\rm Amp}^{(2)}_{NLO}(\boldsymbol{q}) \hspace*{-5pt} &\simeq \hspace*{-5pt} &-\int{d^3\ell\over
(2\pi)^3} {4\pi Gm_1m_2\over |\boldsymbol{p}_f-\boldsymbol{\ell}|^2}{1\over
{p_0^2\over 2m_r}-{\ell^2\over 2m_r} + i \epsilon} {4\pi
Gm_1m_2\over
|\boldsymbol{\ell}-\boldsymbol{p}_i|^2}\nonumber\\
&& \times \left[1+{(p_0^2+\ell^2)\over m_1m_2}\left(\frac{1}{4} + \frac{7}{4} \frac{m_1 m_2}{m_r^2}\right) \right]\nonumber\\
&\simeq \hspace*{-5pt} & - i \hspace*{0.6pt} 4 \pi G^2 m_1^2 m_2^2 \hspace*{1pt} {L\over q^2} {m_r\over p_0}
  + \frac{G^2 m_1^2 m_2^2}{m_1 + m_2} \left(1 + \frac{7 (m_1 + m_2)^2}{m_1 m_2} \right)S . \nonumber \\ \label{eq:jk}
\end{eqnarray}
Subtracting this iterated amplitude, which includes all corrections to NLO, from the scattering
amplitude ${}^0 \! {\cal M}^{(2)}_{tot}(\boldsymbol{q})$ of Eq. (\ref{eq_ampLO_00})
we find the second order potential
\begin{eqnarray}
{}^0V_{NLO}^{(2)}(\boldsymbol r) \hspace*{-5pt} &=\hspace*{-5pt}&-\int{d^3q\over
(2\pi)^3}e^{-i\boldsymbol{q}\cdot\boldsymbol{r}}\left[{}^0 \! {\cal
M}^{(2)}_{tot}(\boldsymbol q)-{}^0{\rm Amp}_{NLO}^{(2)}(\boldsymbol q)\right]\nonumber\\
&=\hspace*{-5pt}&\int{d^3q\over (2\pi)^3} e^{-i\boldsymbol{q}\cdot\boldsymbol{r}} \, G^2 m_1 m_2
\left[ \left((m_1 + m_2) + {m_1m_2\over m_1+m_2}\right) S + {41\over 5}L\right]\nonumber\\
&=\hspace*{-5pt}& \left(1 + \frac{m_1 m_2}{(m_1 + m_2)^2}\right) \frac{G^2 m_1 m_2 (m_1 + m_2)}{2 r^2}
  - {41 G^2 m_1 m_2 \hbar \over 10 \pi r^3} \label{eq:sf}
\end{eqnarray}
and observe that now the classical component agrees with the $\mathcal O(G^2)$ EIH potential
of Eq. (\ref{eq_eihHRV2}).

Thus we have shown that if we consistently take into account the
$v^2$ and $Gm/r$ corrections beyond Newtonian physics we reproduce
the EIH Lagrangian in a certain gauge. From the resulting EIH
Lagrangian we could evaluate observables such as the precession of
the perihelion of Mercury which must clearly be independent of
gauge. The inclusion of the $v^2$ corrections is required since
the equations of motion can be used to describe bound states where
$v^2 \sim Gm/r$ by the virial theorem.

However, our methods are clearly clumsy for the calculation of classical observables. Recently,
Goldberger and Rothstein have developed an effective field theory of gravity which is optimized
for calculating classical observables of bound states called NRGR\cite{gba},\cite{gbb},\cite{gbc},\cite{gbd},\cite{koa},\cite{kob}
Here the external particles are static sources so that no loops are to be calculated in their theory
when calculating classical observables since the only propagating particles present
are gravitons which are massless and thus the loop expansion in NRGR corresponds to an
expansion in $\hbar$. In the NRGR framework the spin-dependent classical equations of motion were
calculated recently to NLO by Porto and Rothstein\cite{poa},\cite{pob},\cite{poc},\cite{pod}
so that we will not continue here to evaluate the corresponding spin-dependent classical
potentials consistently taking into account all relativistic $\mathcal O(v^2)$ effects in the iteration.

\end{document}